%% file: main.tex
\documentclass[letterpaper, 10 pt, conference]{ieeeconf}

\usepackage{epsfig} % for postscript graphics files
\usepackage{amsmath} % assumes amsmath package installed
\usepackage{amssymb}  % assumes amsmath package installed
\usepackage{graphicx}
\usepackage{upgreek}
\usepackage{caption}
\usepackage{color}
\usepackage{subcaption}
\usepackage[utf8]{inputenc}
\usepackage{tikz}
\usepackage{esint}
\usepackage{accents}
\usepackage{hyperref}
\usepackage{enumerate}

\graphicspath{ {./Images/} }

\DeclareMathOperator*{\argmax}{arg\,max}

\definecolor{matred}{rgb}{0.85,0.325,0.098}
\definecolor{matpurple}{rgb}{0.494,0.184,0.556}

\input{my_shortcuts}

\title{Optimal control of a swimming robot \linebreak based on Purcell's microswimmer model}
\author{Noam Berkovich Lahav, Oren Wiezel and Yizhar Or}
\begin{document}

\maketitle

 \begin{abstract}
 Purcell's swimmer is a well-known planar model of a swimming microorganism, governed by low Reynolds number hydrodynamics, which is comprised of three rigid links connected by actuated rotary joints. This model has been analyzed as a robotic locomotion system governed by first-order nonlinear dynamics with a periodic input (gait) of the two joint angles. In this work, we present a robotic macro-scale realization of this three-link swimmer moving in a highly viscous fluid. We propose a simple variant of Purcell's theoretical model with non-slender links and a central rigid sphere which represents the added drag of the robot's central flotation block, and calibrate the model's parameters to fit experimental measurements. Next, we apply optimal control formulation based on Pontryagin's Maximum Principle (PMP) in order to find optimal gaits that maximize the displacement per cycle under bounds on the joint angles. Employing a differential geometric method that transforms the problem to area integral enclosed by the gait trajectory in the plane of joint angles, enables visual interpretation which explains topological changes in displacement-optimal gaits upon varying the bound on the joint angles. We then apply PMP formulation to the problem of maximizing Lighthill's energy efficiency in order to obtain a boundary value problem (BVP) whose solution gives efficiency-optimal gaits for Purcell's swimmer model, as well as its variant with a central sphere. Finally, we utilize numerical methods such as parameterizing the input gait as a truncated Fourier series, as well as GPOPS-II solver, to produce sufficient initial guess values for solving the BVPs and obtaining efficiency-optimal gaits.
 \end{abstract}

\section{Introduction}
The fascinating motion of swimming microorganisms is governed by low Reynolds number ($Re$) hydrodynamics, where viscous drag forces dominate while inertial effects are negligible \cite{happelbrenner_book}. A classic planar model of a swimming microorganism is Purcell's swimmer \cite{purcell1977life}, comprised of three rigid links connected by two actuated rotary joints. A concrete mathematical formulation of this swimmer’s dynamics has been provided in \cite{becker2003self}, where resistive force theory \cite{batchelor1970slender} was utilized to approximate the viscous drag forces acting on slender links.
Due to its quasistatic motion, the swimmer is governed by first-order dynamics of a nonlinear control system, where the actuation inputs are joint angles, often assumed to be prescribed in time-periodic profiles called gaits.
Similar models have been widely investigated in the literature as robotic locomotion systems \cite{kelly&murray95,Ostrowski98}, utilizing geometric methods for studying the motion under gaits that satisfy certain symmetries. 
The work \cite{wiezel2016optimization} studied Purcell's swimmer while assuming sinusoidal gaits, and applied asymptotic analysis in order to obtain explicit expressions for the swimmer's displacement per cycle $X$. 
Importantly, $X$ is not monotonic in the gait's amplitude \cite{becker2003self}, and an analytic approximation of the optimal amplitude that maximizes $X$ was obtained in \cite{wiezel2016optimization} using perturbation analysis. The work \cite{wiezel2016optimization} also obtained asymptotic expression for Lighthill's energy efficiency $\xi$ \cite{lighthill1975mathematica} of circular and square-shaped gaits, and found that $\xi$ is also maximized for certain optimal amplitudes of joint angles.

There have been very few experimental demonstrations of Purcell's three-link swimmer \cite{kumar2011robo,kadam2017trajectory,chan2009bio}. The work \cite{kosa_tro07} presented a macro-scale swimmer in glycerol ($Re\sim 1$) using sinusoidal inputs for piezoelectric actuators mounted along an elastic beam. The work \cite{Gutman2015Symmetries} presented a macro-scale three-link swimmer in silicone fluid ($Re \sim 0.4$) that demonstrated motion under gaits with certain symmetries. The swimmer in \cite{Gutman2015Symmetries} contained a large central block of flotation foam for carrying the motor and controller, which produced added viscous drag. This effect was not included in Purcell’s original model, and thus \cite{Gutman2015Symmetries} obtained only qualitative agreement with theoretical analysis, while focusing on gaits with different symmetries.  In addition, both robots in \cite{kosa_tro07,Gutman2015Symmetries}  were tethered to power supply cable, adding non-negligible external tension forces.
In the micro-nano scales, several artificial miniature swimmers were presented. Some of them involve passive elastic structures or hinges \cite{gao2012cargo,jang2015undulatory} while others contain rigid helices \cite{ghosh.nanoswim2009,PeyerNelson2013}. Actuation methods included external magnetic field \cite{dreyfus2005microscopic,HelicalOrNelsonChapnik2021}, internal chemical reaction \cite{gao2011hybrid}, and even a bio-hybrid microswimmer with electric activation of a particle attached to a living flagellated bacterium \cite{carlsen2014bio}. In the large-$Re$ regime of inertial swimming \cite{miloh1993self}, several experimental prototypes of multi-link snake-like robotic swimmers have been presented \cite{kelasidi2015experimental,mcisaac2002experimental,porez2014improved,virozub2019planar}. These robots utilized controlled servo motors to prescribe sinusoidal inputs for joint angles, but did not analyze the effect of angle’s amplitude.

While most work cited above focused on sinusoidal or square-wave inputs, an important question is to find optimal gaits for Purcell’s swimmer which can take \textit{any} time-periodic function. A first locally-optimal gait trajectory that maximize displacement has been obtained in \cite{tam2007optimal} using direct numerical optimization of coefficients of truncated Fourier series. Moreover, the follow-up comment in \cite{avron2008comment} suggested using a self-intersecting dumbbell-shaped gait to increase the net displacement further. Later, the problem was optimized analytically for the first time in \cite{giraldi2015optimal,wiezel2016using}, using optimal control theory \cite{bryson1975applied,ben_asher_2010optimal} which is based on variational calculus and Pontryagin’s Maximum Principle (PMP) \cite{pontryagin1987mathematical}. This work was extended recently in \cite{wiezel2023geometric} to incorporate bounds on joint angles into the variational formulation. The problem has also been analyzed using differential-geometric method \cite{shapere1989geometry,hatton2013tro}, transforming the displacement to area integral enclosed by loops of the gait trajectories, for a height function which is obtained by curvature of connections for optimized choice of minimum-rotation body-coordinates \cite{hatton2010optimizing}. 
This enables visual interpretation, and the displacement-optimal gait of \cite{tam2007optimal} was obtained as a zero-level closed curve of the height function, and compared with the PMP analysis in \cite{wiezel2023geometric}. However, such a geometric analysis has not yet been applied to an experimentally validated Purcell-type swimmer model. 

Considering maximization of Lighthill’s energy efficiency $\xi$, an efficiency-optimal gait for Purcell’s swimmer was obtained numerically in \cite{tam2007optimal} using a similar approach of optimizing coefficients of Fourier series coefficients. The same gait was also obtained in \cite{ramasamy2016soap,ramasamy2019geometry} using a different scheme of “soap-bubble optimization” with added geometric interpretation. Nonetheless, analyzing the optimization of $\xi$ for Purcell-type swimmers using PMP formulation has not yet been considered in the literature.

In this work, we present experimental results of our macro-scale untethered three-link robot moving in silicone fluid. Using motion tracking measurements under sinusoidal gaits, we obtain an optimal amplitude which maximizes the displacement per cycle. In order to improve quantitative agreement between theory and experiment, we consider a variant of Purcell’s model with non-slender links and a central rigid sphere which represents the added drag of the robot’s central flotation block. We then utilize numerical simulations in order to calibrate the model’s parameters to fit experimental measurements of our robotic swimmer.  Next, we apply the optimal control formulation from \cite{wiezel2023geometric} in order to find optimal periodic gaits for this swimmer model that maximize the displacement per cycle under bounds on the of joint angles. This results in displacement-optimal gaits which differ significantly in their qualitative shape from those of Purcell’s swimmer original model. Employing the differential-geometric analysis from \cite{wiezel2023geometric,hatton2013tro} enables visual interpretation which explains topological changes in displacement-optimal gaits upon varying the joint angle bounds. Finally, we analyze the problem of maximizing Lighthill’s energy efficiency using PMP formulation for the first time. This results in a two-point boundary value problem whose solution gives efficiency-optimal gaits for Purcell’s swimmer as well as its variant with a central sphere. This problem is sensitive to the choice of the initial guess values. Therefore, we employ numerical tools in order to obtain well-informed initial guesses, such as parameterizing the input gait as a truncated Fourier series, as well as \textit{GPOPS-II} solver \cite{GPOPS2Patterson2014}, an optimal control solver based on hp-adaptive Gaussian quadrature collocation. Using this method, we reproduce the efficiency-optimal gait for Purcell's swimmer model found in \cite{tam2007optimal}, and further discover a new large-amplitude efficiency-optimal gait with even greater efficiency. In addition, for the modified robotic swimmer model, we find two distinct efficiency-optimal gaits, one at moderate amplitudes and one at large amplitudes which exhibit higher efficiency. These gaits differ substantially in their shape from their counterparts in Purcell's swimmer model. All results were also reproduced numerically using \textit{GPOPS-II}.
 
\section{Purcell's microswimmer model and the robotic prototype}
In this section, we review Purcell's swimmer and the formulation of its dynamic equations of motion. Then, we introduce an experimental low-$Re$ three-link macro-scale robotic swimmer prototype and modify Purcell's theoretical model to better fit its dynamics.

\subsection{Purcell's low-\texorpdfstring{$Re$}{Re} three-link swimmer model}
\label{sec:purcell_model}
Purcell's swimmer dynamic equations of motion were first formulated by Becker et al \cite{becker2003self}. This three-link planar swimmer is submerged in an unbounded fluid domain whose motion is governed by Stokes equations \cite{happelbrenner_book}. The swimmer model consists of three thin rigid links with lengths $l_0,l_1,l_2$, where $l_1=l_2$. The links are connected by two rotary joints whose angles are denoted by $\phi_1$ and $\phi_2$ (see Fig. \ref{fig:Purcell_swimmer}). The shape of the swimmer is denoted by the shape variables $\Phi=(\phi_1,\phi_2)^T$. It is assumed that the swimmer's motion is confined to a plane. The swimmer's body motion is defined as the planar position of the middle link's center $(x,y)^T$ and its orientation angle $\theta$; these constitute the swimmer's body variables $\vecq_b=(x,y,\theta)^T$. The velocity of the central link in an inertial frame is denoted by $\dot \vecq_b=(\dot x, \dot y, \dot\theta)^T$. The velocity of the $i$th link is described by the linear velocity of its center $\vecv_i=(\dot x_i,\dot y_i)^T$ and the link's angular velocity $\omega_i$, which are augmented in the vector $\vecV_i=(\vecv_i,\omega_i)^T\in\mathbb{R}^3$. The swimmer is submerged in an unbounded fluid domain whose motion is governed by Stokes equations \cite{happelbrenner_book}.

Resistive force theory for slender links \cite{gray1955propulsion} states that the viscous drag force $\vecf_i$  and torque $m_i$  on the $i$th slender link under planar motion are proportional to its linear and angular velocities.
\begin{equation}
\begin{array}{c}
\vecf_i=-c_t l_i(\vecv_i\cdot\vect_i)\vect_i-c_n l_i(\vecv_i\cdot\vecn_i)\vecn_i \\[12pt]
m_i=-\dfrac{1}{12} c_n l_i^3 \omega_i,
\end{array}
\label{eq:RFT}
\end{equation}
Where $\vect_i=(\cos\alpha_i,\sin\alpha_i)^T$ and $\vecn_i=(-\sin\alpha_i,\cos\alpha_i)^T$ are unit vectors in the axial and normal direction of the $i$th link respectively, i.e. $\alpha_0=\theta, \alpha_1=\theta+\phi_1, \alpha_2=\theta+\pi-\phi_2$, and $c_n, c_t$ are the resistance coefficients for the normal and axial directions.

Equation \eqref{eq:RFT} can be rewritten as a relation between the generalized force on each link $\vecF_i=(\vecf_i,m_i)^T$ and its generalized velocity $\vecV_i$:
\begin{equation}
\vecF_i=-\vecR_i(\Phi,\theta)\vecV_i
\label{eq:Fi}
\end{equation}
Where the resistance matrix of the $i$th link $\vecR_i$ can be written as:
\small
\begin{equation}
\vecR_i=c_t^{(i)}l_i
\begingroup % keep the change local
\setlength\arraycolsep{1pt}
\begin{bmatrix}
1\!+\!(\chi\!-\!1)\sin^2\alpha_i & (1\!-\!\chi)\sin\alpha_i\cos\alpha_i & 0\\
(1\!-\!\chi)\sin\alpha_i\cos\alpha_i & 1\!+\!(\chi\!-\!1)\cos^2\alpha_i & 0\\
0 & 0 & \frac{\chi}{12}l_i^2
\end{bmatrix}
\endgroup
\label{eq:Ri}
\end{equation}
\normalsize
Where $\chi\!=\!c_n/c_t$. In the limit of an infinitely slender rod, the resistance coefficients in (\ref{eq:RFT}) are given as $c_t=0.5c_n=\frac{2\pi\mu}{log(l_i/a)}$, where $\mu$ is the fluid's viscosity and $a$ is the rod's cross-section radius \cite{cox1970}. That is, for a slender rod it is assumed that $\chi=2$.

Using the swimmer's kinematic relation to express $\vecv_i$ in terms of $\vecq_b, \dot\vecq_b,\Phi,\dot\Phi$, the net hydrodynamic force and torque acting on the swimmer's body $\vecF_b$ can be derived (see \cite{wiezel2016optimization}).
Assuming quasi-static motion, the swimmer is in static equilibrium $\vecF_b=0$ and the swimmer's dynamic equations of motion are obtained as a linear relation between the joints' velocities and the swimmer's body velocities:
\begin{equation}
\dot\vecq_b=\vecG(\Phi,\theta)\dot\Phi
\label{eq:full_connection}
\end{equation}
This means that the swimmer's body motion is time-invariant, i.e. independent of its shape change rate and depends only on the trajectory of $\Phi$ ($d\boldsymbol{q}=\vecG(\Phi,\theta)d\Phi$) \cite{purcell1977life}.   

As in previous studies on Purcell's swimmer \cite{wiezel2016using,wiezel2023geometric}, we focus on the case where the input joint angles $\Phi(t)$ are periodic, thereby generating a gait. Our objective is to identify two types of optimal gaits. The first maximizes the net displacement achieved over a single cycle, while the second maximizes Lighthill's energetic efficiency, explained as follows. 

In order to examine the energetic efficiency of the swimmer's motion, we require expressions for the mechanical energy expenditure, which is the time integral of the mechanical power $E(t) = \int p(t) dt$.
As shown in \cite{wiezel2016optimization}, the mechanical power can be expressed as:
\begin{equation}
p(t) = \dot\Phi^T \vecW (\Phi) \dot\Phi
\label{eq:power}
\end{equation}
where $\vecW(\Phi)>0$ is a symmetric positive definite power matrix, the formulation of which is given explicitly in \cite{wiezel2016optimization}. 
Lighthill's efficiency $\xi$ is defined as the ratio between the power required to drag the swimmer by an external force at its average velocity $\bar{V}$ and the average power $\bar{P}$ expended along the gait \cite{lighthill1975mathematica}:

\begin{equation}
\xi\!=\! \frac{c_tl\bar{V}^2}{\bar{P}}
\label{eq:LHE_DEF}
\end{equation}

\subsection{A robotic low-\texorpdfstring{$Re$}{Re} three-link swimmer prototype}
\label{sec:sphere_model}

\begin{figure}[h]
\centering
\begin{subfigure}[c]{0.23\textwidth}
\includegraphics[width=\textwidth]{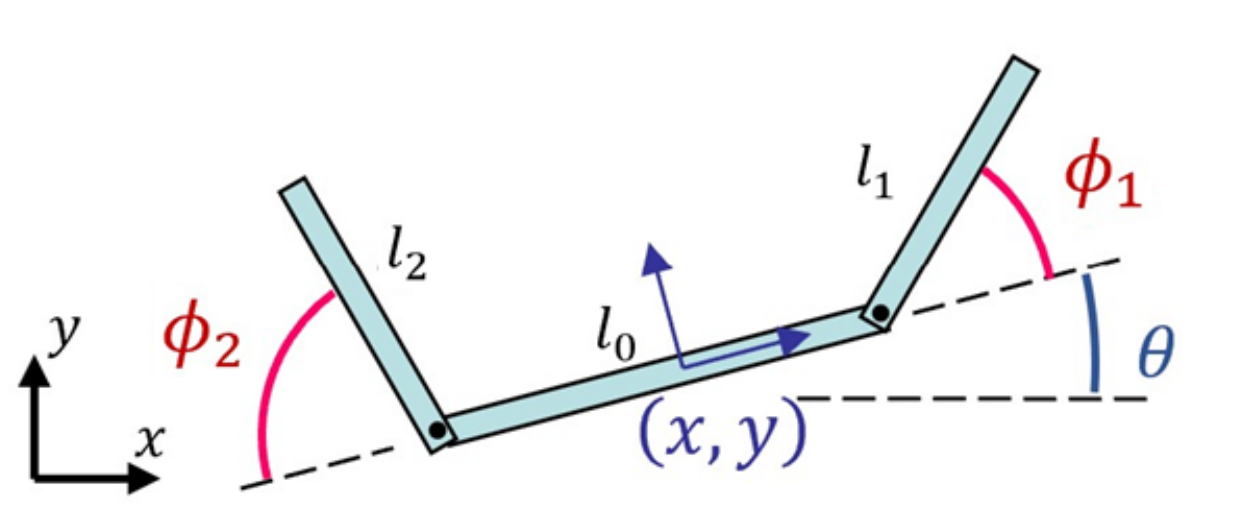}
\caption{}
\label{fig:Purcell_swimmer}
\end{subfigure}
\begin{subfigure}[c]{0.23\textwidth}
\includegraphics[width=\textwidth]{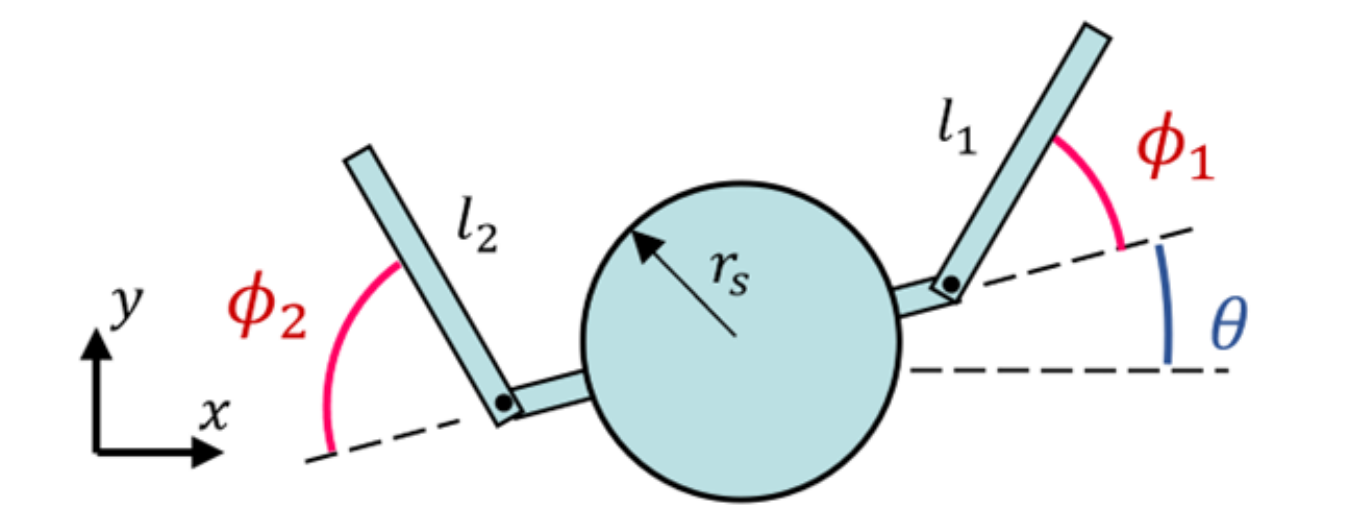}
\caption{}
\label{fig:swimmer_sphere}
\end{subfigure}
\begin{subfigure}[c]{0.23\textwidth}
\includegraphics[width=\textwidth]{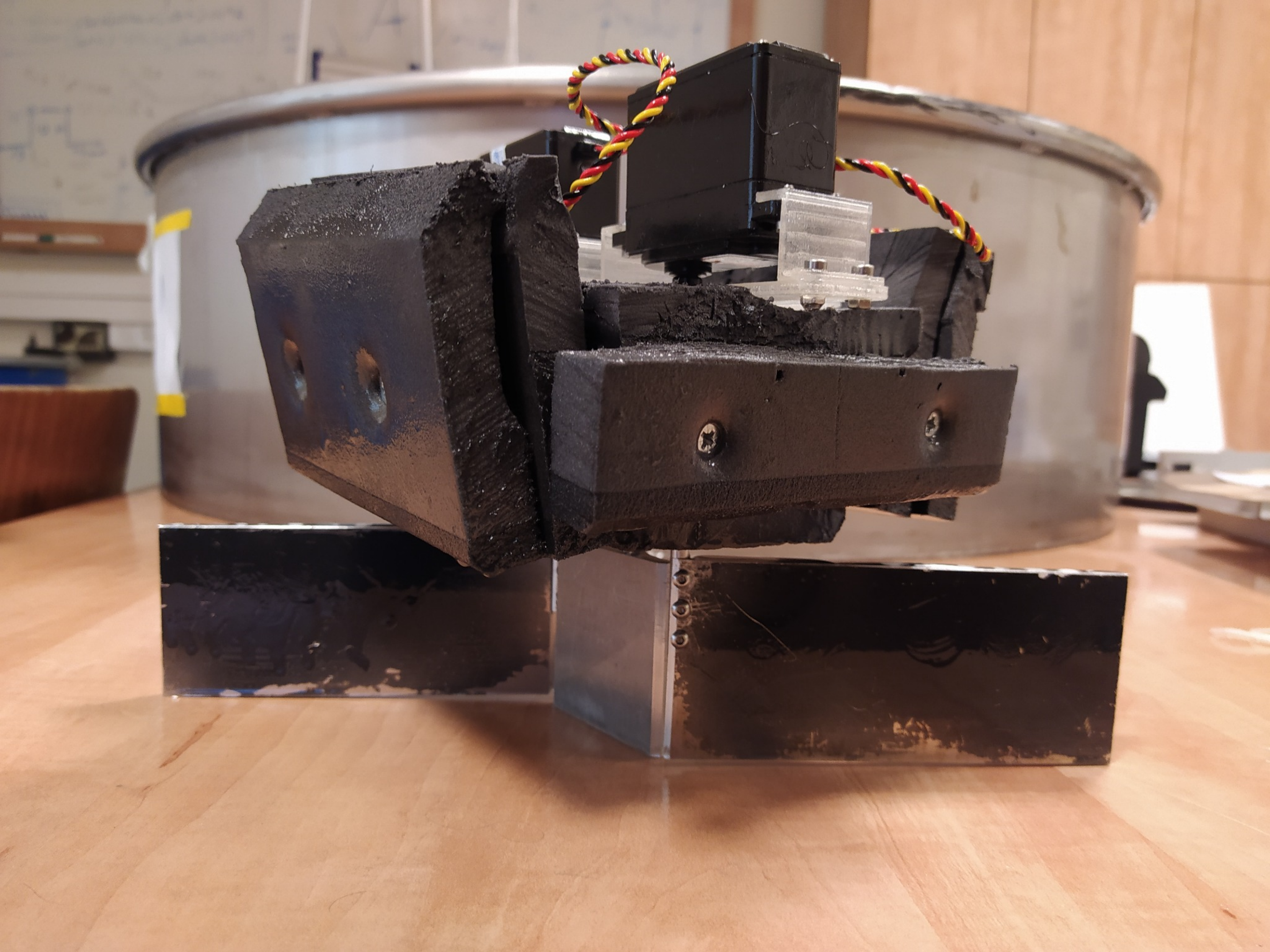}
\caption{}
\label{fig:RobotPrototype}
\end{subfigure}
\begin{subfigure}[c]{0.23\textwidth}
\includegraphics[width=\textwidth]{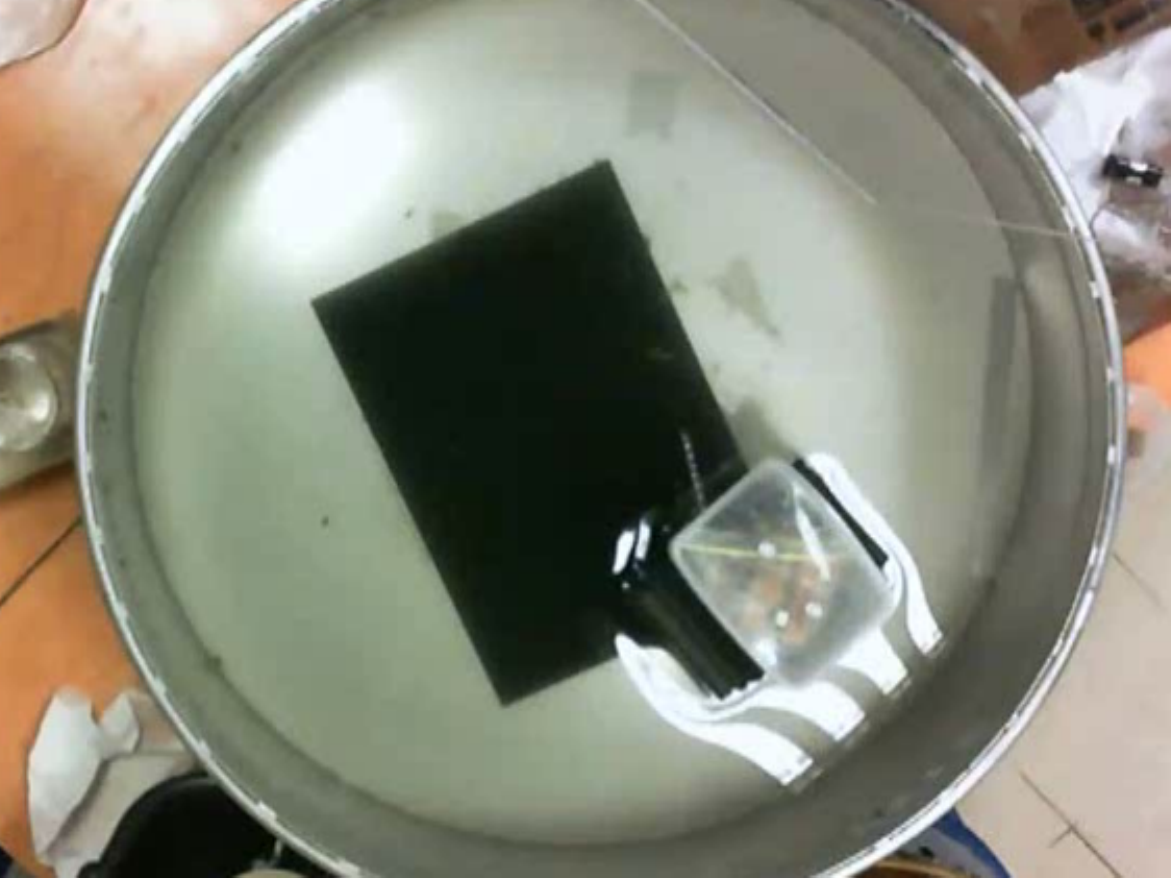}
\caption{}
\label{fig:RobotInTank43}
\end{subfigure}
 \caption{ (a) Model of Purcell's three-link swimmer. \\(b) Model of the three-link swimmer with central sphere. \\(c) The robotic prototype of a low Reynolds number three-link swimmer. (d) The robotic prototype in silicone oil tank.}
 \label{Fig:RobotPhotos}
 \end{figure}

We now present a robotic prototype of our macro-scale three-link swimmer. This robot is an improved, untethered version of the earlier tethered robot in \cite{Gutman2015Symmetries}. 
The robotic swimmer prototype, shown in Fig. \ref{fig:RobotPrototype}, consists of three thin aluminum plates that represent the links.
The thickness of the plates is $2a=0.5 \,[cm]$ and their width in the vertical direction is $b=6 \,[cm]$. The length of the middle plate is $l_0 =7.1\, [cm]$ and of the side plates is $l_1=l_2=13.7 \,[cm]$. 
The links are connected by joints which are actuated by two servo motors (HITEC HS-785HB) that are mounted on top of the links. The motors were controlled using an ARDUINO NANO controller with an HC-05 Bluetooth module for receiving commands. The motors, controller, and Bluetooth module as well as four AA batteries, were placed on top of the center link on a foam floatation cell while the aluminum plates are immersed in the viscous fluid. 
The role of the floatation cell is to maintain planar motion of the swimmer in the horizontal plane, as well as holding the motors and electronics outside of the fluid. The fluid is a highly viscous silicone oil with kinematic viscosity of $\nu=60,000\,cSt$ and specific gravity of $0.976$. Characteristic angular velocity of the joint is $\omega=0.24\,[rad/sec]$. 
The characteristic velocity is taken as the mean velocity of the side link due to joint rotation i.e. $v=0.5l_2\omega=0.0168\,[m/s]$. 
Thus, the characteristic Reynolds number for the experiments is estimated as $Re=vl/\nu\approx0.039$. This verifies that the motion is under the regime of low Reynolds number hydrodynamics, and the inertial effects are indeed negligible for this macro-robot due to the high viscosity of the fluid and the slow pace of motion.
The experiments were conducted in a rounded container with a diameter of $1\,[m]$ (Fig. \ref{fig:RobotInTank43}). An infrared Optitrack Flex V:100 camera was used for measurements of the swimmer's planar motion. The camera tracks two reflectors attached to the motors in a sampling rate of $100\,[Hz]$, and the measurements were processed using \textsc{Matlab}.

\begin{figure}[ht]
\centering
\begin{subfigure}[c]{0.23\textwidth}
\includegraphics[width=\textwidth]{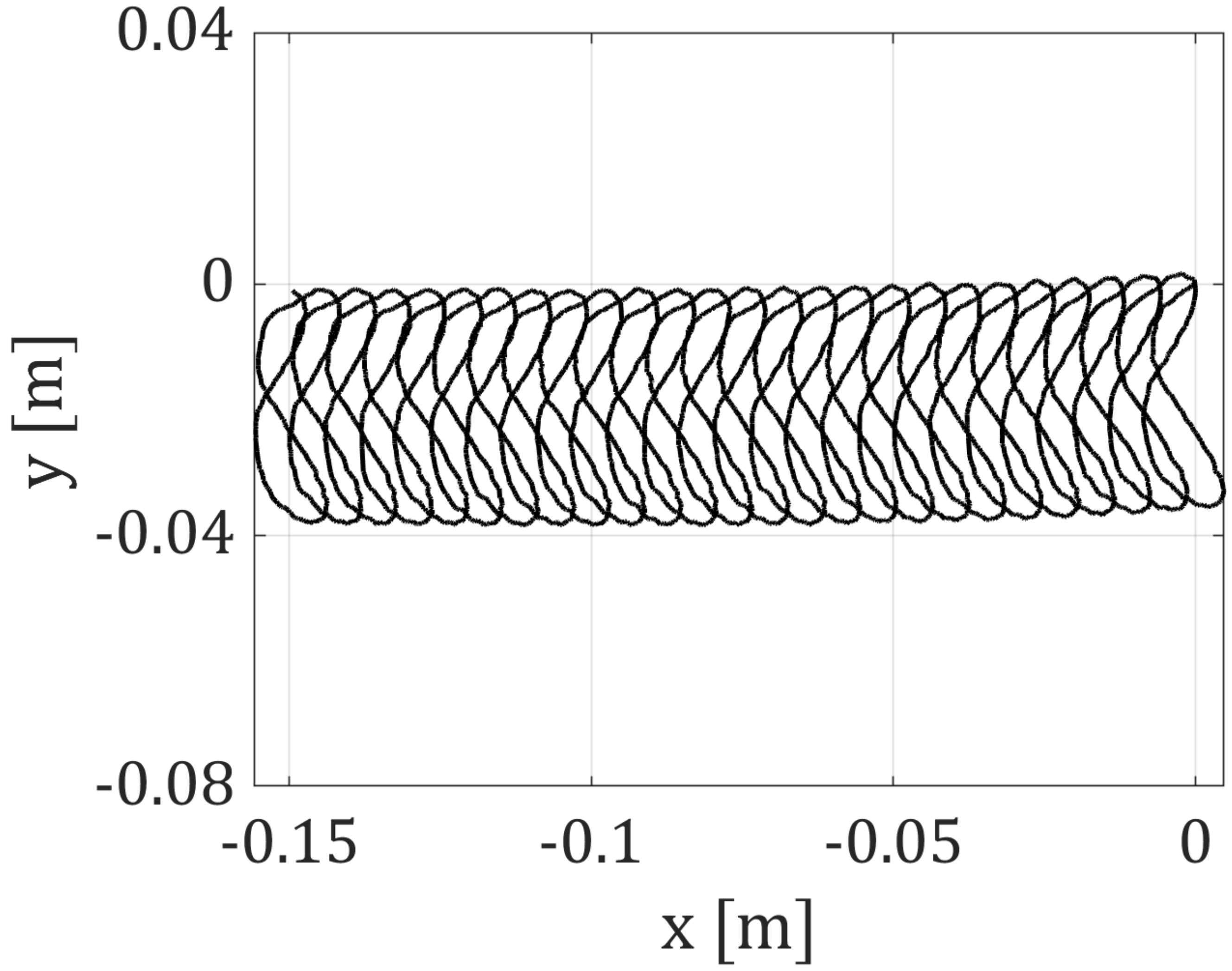}
\caption{}
\label{fig:y_x_exp_70_deg}
\end{subfigure}
\begin{subfigure}[c]{0.245\textwidth}
\includegraphics[width=\textwidth]{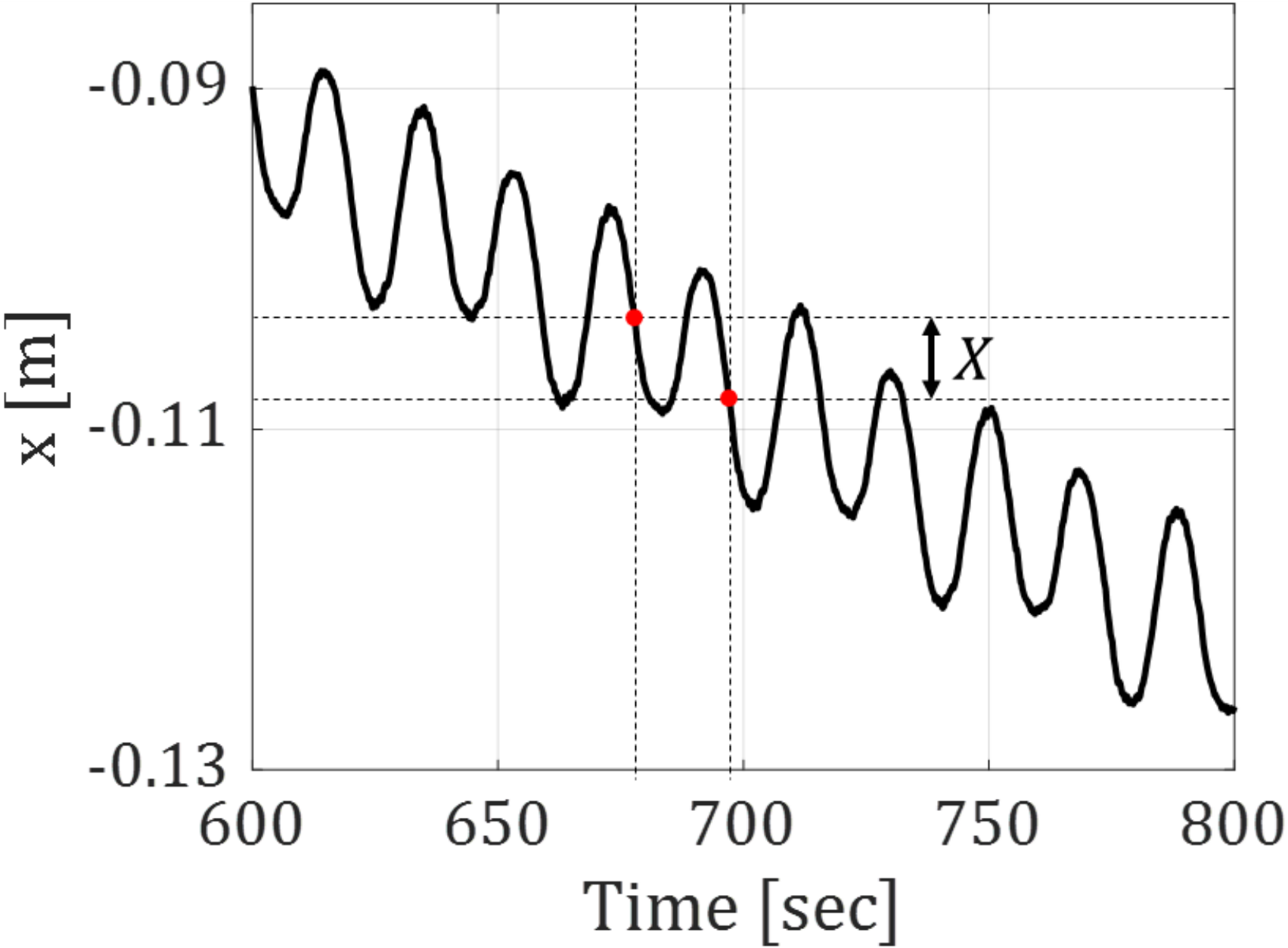} 
\caption{}
\label{fig:x_time_exp_70_deg}
\end{subfigure}
\begin{subfigure}[c]{0.235\textwidth}
\includegraphics[width=\textwidth]{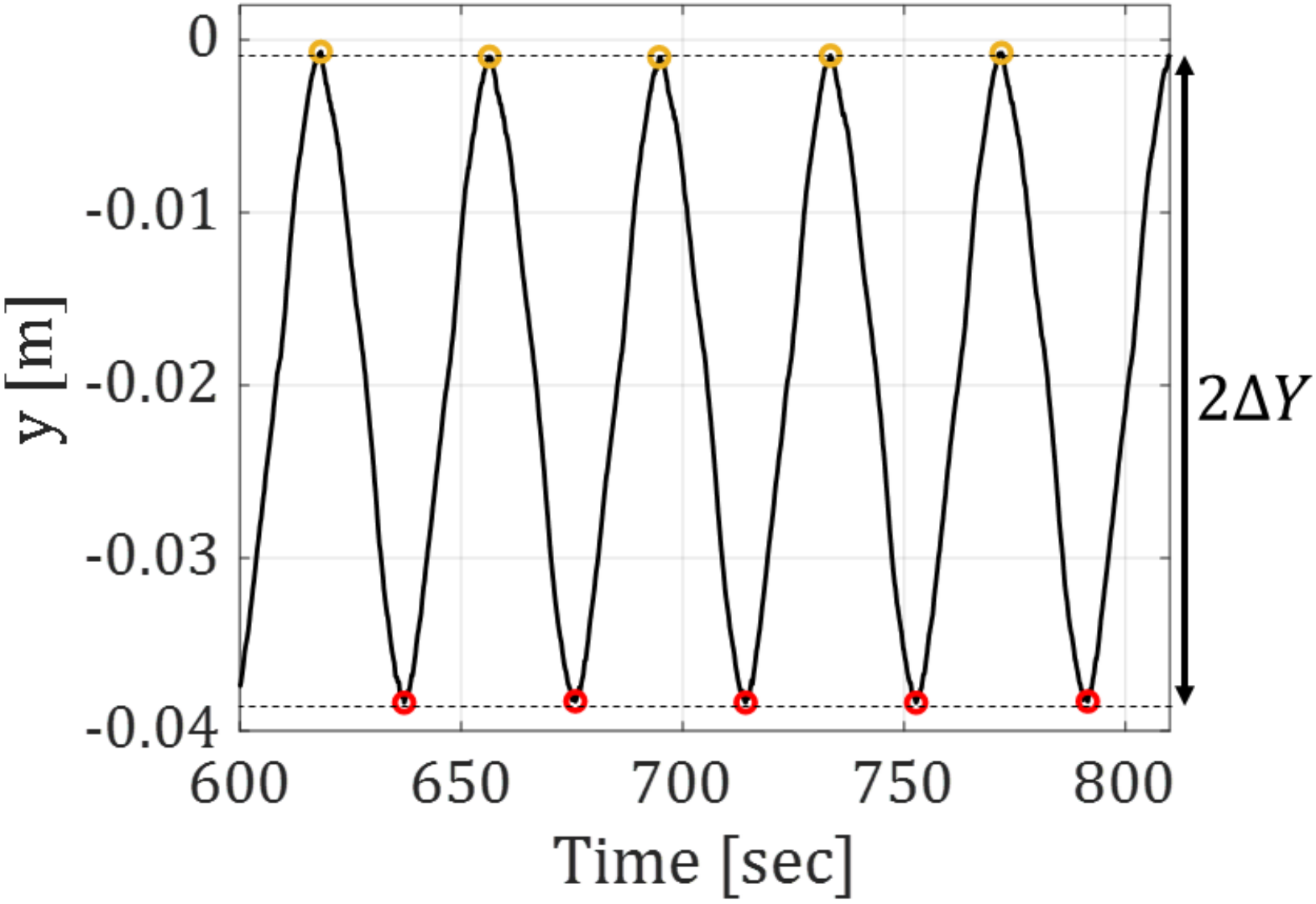}
\caption{}
\label{fig:y_time_exp_70_deg}
\end{subfigure}
\begin{subfigure}[c]{0.235\textwidth}
\includegraphics[width=\textwidth]{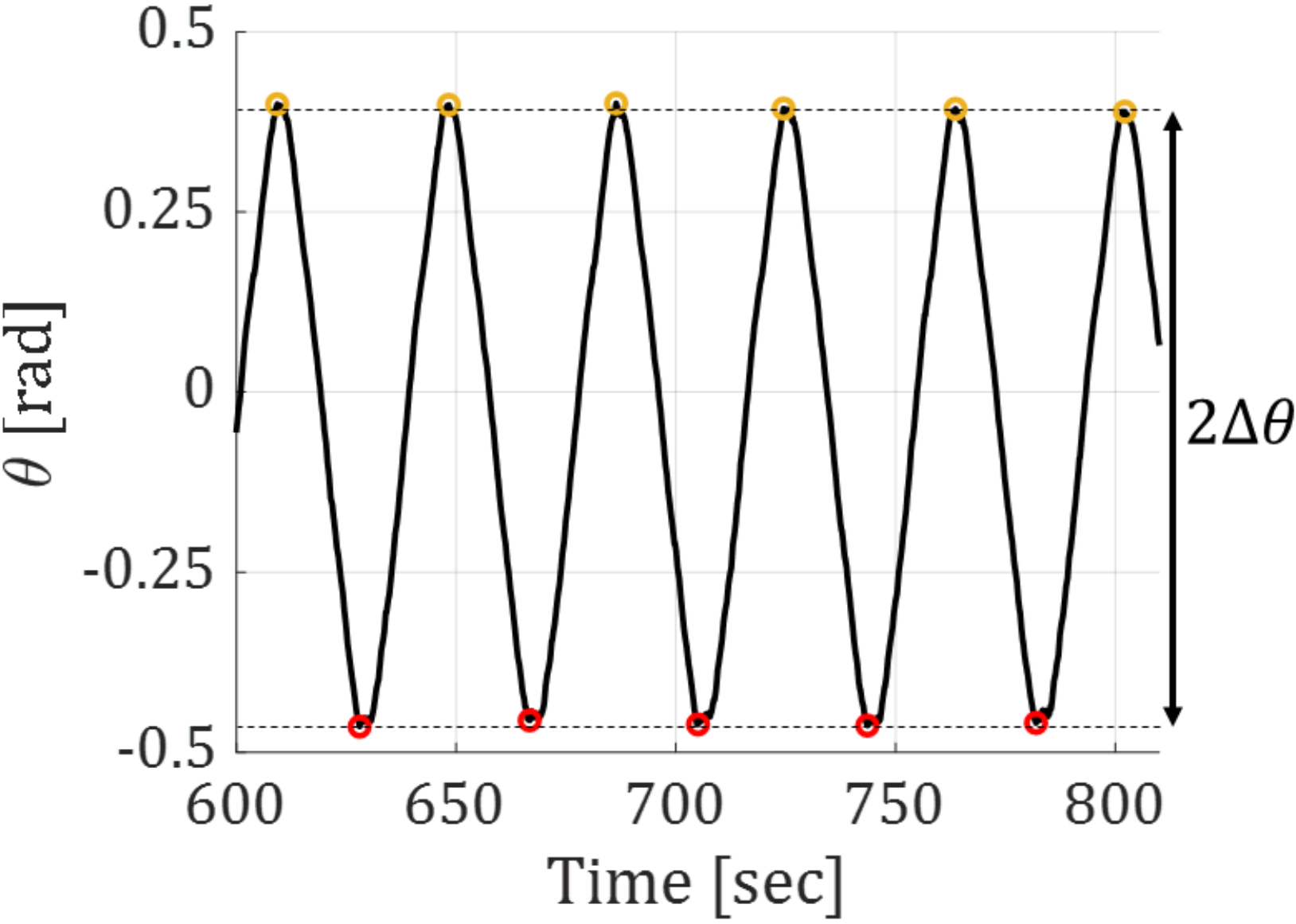} 
\caption{}
\label{fig:theta_time_exp_70_deg}
\end{subfigure}
 \caption{Results of an experiment on the robotic swimmer for a circular gait input with an amplitude of $\varepsilon=70^\circ$: \\(a) Trajectory of the central link’s midpoint in x-y plane, \\(b) $x(t)$, (c) $y(t)$, (d) $\theta(t)$.}
 \label{Fig:Exp_Res}
 \end{figure}

In the experiments presented here, the shape changes were defined as harmonic inputs: $\phi_1(t)\!=\!\varepsilon\cos(\omega t-\psi/2)$ and $\phi_2(t)\!=\!\varepsilon\cos(\omega t+\psi/2)$,
with a period time of $T=2\pi/\omega\approx30 [sec]$. A set of experiments with varying amplitudes was performed for a circular gait (phase difference $\psi=\pi/2$). Multiple periods of each gait were performed  ($\approx 20$ periods). The results of an example experiment for a circular gait input with an amplitude of $\varepsilon=70\,[deg]$ are shown in Fig. \ref{Fig:Exp_Res}.       

The $x$ direction was taken as the orientation of the center link at the time when the swimmer is in a symmetric initial state ($\phi_1=\phi_2>0$) and the mean displacement was taken as the displacement per period ($X$). The results show very small net displacement in $y$ direction and net rotation over a period, in agreement with the theoretical prediction due to the gait's symmetry \cite{Gutman2015Symmetries}. Since this net change over a period is close to zero, for the $y$ direction and the rotation $\theta$, we considered the amplitude of the oscillations during a period, denoted $\Delta Y$ and $\Delta \theta$, as illustrated in Fig. \ref{Fig:Exp_Res}. The amplitude was calculated as half the difference between the maximal and minimal value in a period. These values further characterize the swimmer's motion and will be used later, in \ref{sec:modified_model}, to improve the agreement between the theoretical model and the actual robotic swimmer.  

\begin{figure}[h]
\centering
 \begin{subfigure}[c]{0.185\textwidth}
\includegraphics[width=\textwidth]{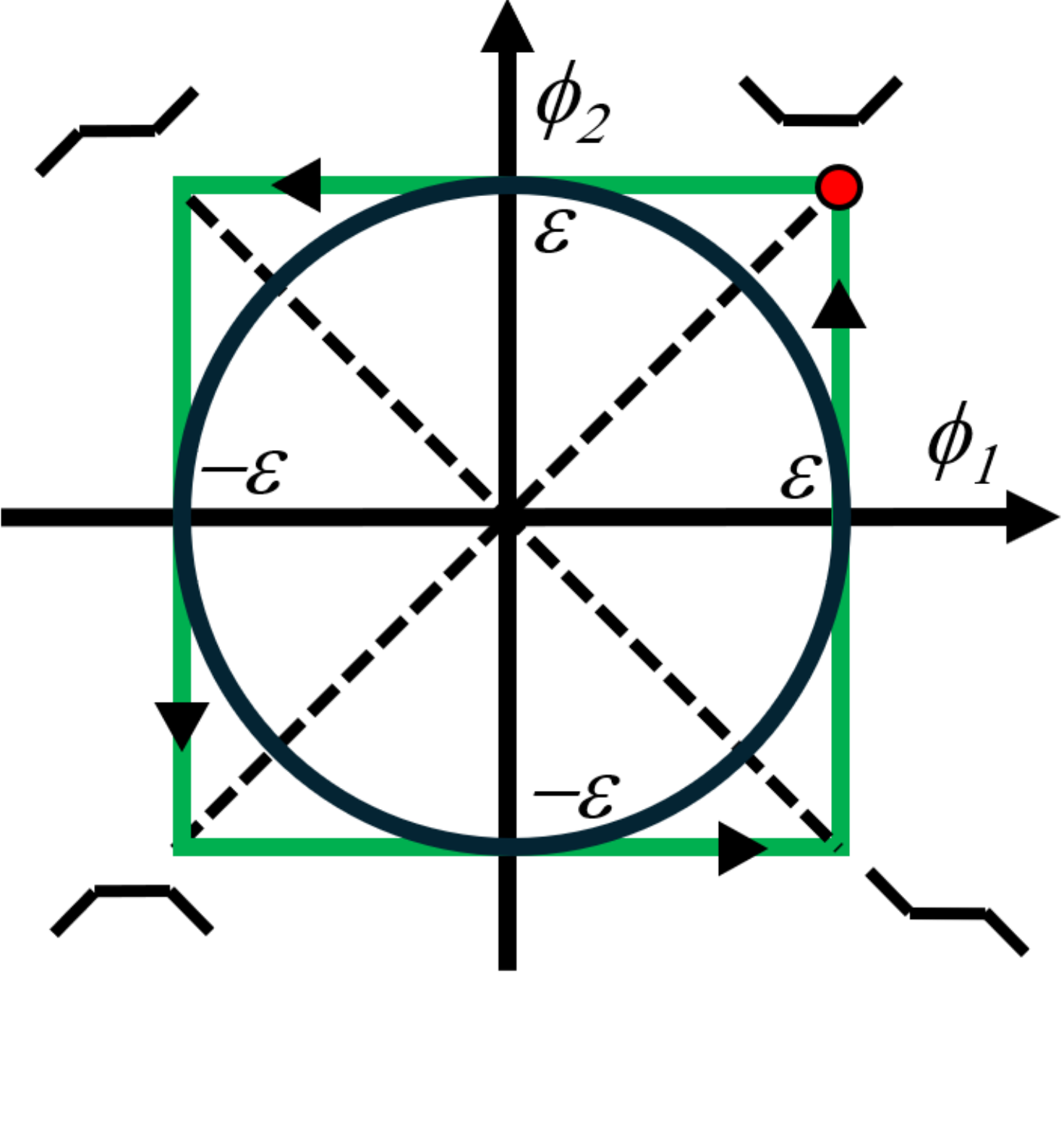}
\caption{}
\label{fig:Circ_and_Squ_Gait}
\end{subfigure}
\begin{subfigure}[c]{0.235\textwidth}
\includegraphics[width=\textwidth]{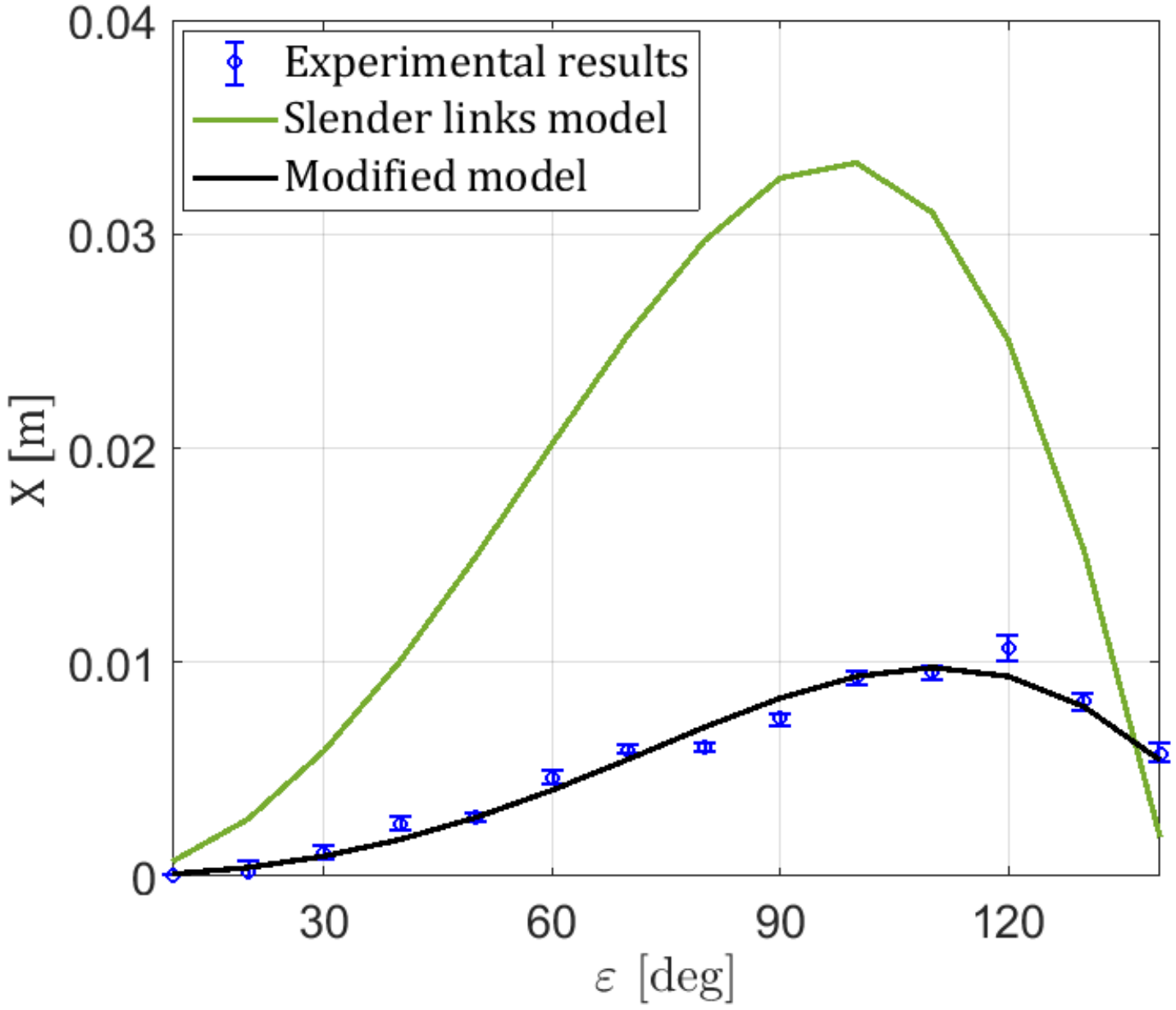}
\caption{}
\label{Fig:exp_amp}
\end{subfigure}
\caption{(a) Gait representation in joint angle plane for circular and square-shaped gaits. (b) Experimental results vs simulation of net displacement in $x$ direction over a period with and without model correction, for circular gaits of varying amplitude $\varepsilon$.}
\label{fig:Circ_Squ_and_Exp_Results_Disp}
\end{figure}

Results of the set of circular gait input experiments, under different stroke amplitudes $\varepsilon$, are shown in Fig. \ref{Fig:exp_amp}. The blue marks in the graph represent the mean net displacement per cycle $X$ as a function of the gait amplitude $\varepsilon$.  The error bars represent standard deviation over periods of each run of the experiment. Importantly, these results demonstrate the existence of an optimal amplitude, where $X$ attains local maximum, verifying the theoretical analysis in \cite{becker2003self} and \cite{wiezel2016optimization}. The green line represents the simulation results for Purcell's swimmer model with slender links as shown in \cite{wiezel2016optimization}, using the same link lengths $l_i$ of the robot. 

As can be seen clearly, there is a significant difference between the simulation and experimental results under the same input of joint angles' gait. The maximal net displacement of the robotic swimmer is more than 3 times smaller than its theoretical counterpart, suggesting unaccounted drag effects. This is mainly due to the key differences between Purcell's theoretical model and the actual robotic prototype. While Purcell's swimmer model assumes three slender links with a cylindrical cross-section, the robot links are made of three thin rectangular plates. Furthermore, the robotic prototype has a large flotation cell at the center that adds significant drag on the swimmer's central body and is not represented in Purcell's model. 

\subsection{The robotic swimmer's modified model}
\label{sec:modified_model}

In order to compensate for the differences between the theoretical model and the robot and improve the model's predictions, we adjusted the theoretical model to represent the dissimilarities. The first adjustment was adding a sphere of radius $r_s$ located at the middle of the central link of the swimmer model, in order to represent the added drag due to the flotation cell (see Fig. \ref{fig:swimmer_sphere}).
The relation between the forces and torques acting on a sphere and its linear and angular velocities are given in \cite{happelbrenner_book,Gutman2016PREOptimizing}, from which the resistance matrix of a sphere can be written:

\begin{equation}
\vecR_{sphere}=\mu\pi
\begin{bmatrix}
6r_s & 0 & 0\\
0 & 6r_s & 0\\
0 & 0 & 8r_s^3
\end{bmatrix}
\label{eq:R_sphere}
\end{equation}

In order to write the modified swimming equation in the form of \eqref{eq:full_connection}, we now only need to add the sphere's resistance to the central link's resistance matrix: $\vecR_0^{new}=\vecR_0+\vecR_{sphere}$, assuming that hydrodynamic interactions between links can be neglected.
 
Additionally, we allow a varying ratio of $\chi$ between normal and axial drag coefficients for the thin links. For finite-length ellipsoidal bodies the ratio is lower than $2$ \cite{berman2013undulatory}. On the other hand, for thin plates, this ratio can be larger than $2$ \cite{jones1958drag}. 

We chose the parameters $r_s$ and $\chi$ by fitting to the experimental results, using \textsc{Matlab}'s \texttt{fmincon} command to minimize the weighted root mean square error function:  
\begin{equation}
RMSE=\sqrt{\sum_{i=1}^{n}\left(w_xq_x^{(i)}+w_yq_y^{(i)}+w_{\theta}q_{\theta}^{(i)}\right)}
\label{eq:RMSE}
\end{equation}
Where $q_x^{(i)}=\left(\frac{X_{sim}^{(i)}-X_{exp}^{(i)}}{\max(X_{exp})}\right)^2$, $q_y^{(i)}=\left(\frac{\Delta Y_{sim}^{(i)}-\Delta Y_{exp}^{(i)}}{\max(\Delta Y_{exp})}\right)^2$, $q_{\theta}^{(i)}=\left(\frac{\Delta\theta_{sim}^{(i)}-\Delta\theta_{exp}^{(i)}}{\max(\Delta\theta_{exp})}\right)^2$ and the data points where evaluated as the average value over several cycles. We choose the weights of the $RMSE$ to be $w_x\!=\!0.5$ and $w_y\!=\!w_{\theta}\!=\!0.25$, such that the dominant fitting criterion will be the swimmer's net displacement $X$. 

The parameter fitting values obtained for the modified model were $r_s=6.92 \, cm$ and $\chi=3.598$. Simulation results of the net displacement $X$ versus circular gait amplitude using the modified model are presented in Fig. \ref{Fig:exp_amp} in the black curve. 
Clearly, fitting these two parameters drastically improves the agreement between the simulations and the experiments compared to the original model with slender links.

\section{Optimal control using Pontryagin's maximum principle}
\label{sec:OCP}
In order to identify gaits that maximize net displacement per cycle or Lighthill's efficiency of the considered swimmer models, we now review the formulation of an optimal control problem (OCP) and employ Pontryagin's maximum principle (PMP) to solve it \cite{bryson1975applied}.  
Below, We outline the general formulation of such a problem, its general solution via PMP and some required OCP extensions. 

\subsection{Formulation of OCP}
We begin with a short description of the optimal control problem and PMP solution for a model with the dynamics as in \eqref{eq:full_connection}. 

Consider a dynamical control system $\dot\vecz=\vecf(\vecz,\vecu)$, where $\vecz\in\mathbb{R}^n$ is the state of the system and $\vecu$ is the control input. We are looking for the optimal control input $\vecu^*(t)$ that maximizes a certain final-time cost function $J(t_f,\vecz_f)$. The following OCP formulation is obtained:
\begin{equation}
(OCP)
	\begin{cases}
	\max J(t_f,\vecz_f) &\ s.t.\\
	\dot\vecz=\vecf(\vecz,\vecu)& \forall t\in [t_0,t_f],\\
	\vecu \in \mathcal{U} & \forall t \in [t_0,t_f],\\
	\end{cases}
\end{equation}
In order to solve this problem, we use Pontryagin's Maximum Principle (PMP). We define the Hamiltonian as:
\begin{equation}
\label{eq:Hamiltonian_Def}
H(\vecz,\vecu,\veclambda)=\veclambda^T \vecf
\end{equation}
where $\veclambda\!\in\!\mathbb{R}^n$ is a vector of costate variables with the dynamics:
\begin{equation}
\label{eq:Costate_Dynamics_Def}
\dot \veclambda = -\frac{\partial H}{\partial \vecz}  
\end{equation}
Pontryagin's Maximum Principle (PMP) states that the optimal control input trajectory $\vecu^*(t)$ with associated state trajectory $\vecz^*(t)$ is the one that maximizes the Hamiltonian:
 \begin{equation}
 \vecu^*(\vecz,\veclambda)=\argmax_{\vecu\in\mathcal{U}} H(\vecz,\vecu,\veclambda).
 \end{equation}
 and the optimal input can be found by solving 
 \begin{equation}
 \vecH_\vecu=\frac{\partial H}{\partial \vecu}=\mathbf{0}
 \label{eq:Hu_general}
 \end{equation}
This Maximum Principle requires solving an ODE with two-point boundary conditions in order to solve the OCP.

\subsection{OCP Extensions}
Some OCPs require additional tools in order to properly define and solve them. This section shortly reviews some of these tools, which are required for our study.

\subsubsection*{Singular arcs}
In problems where the Hamiltonian $H$ is linear in the input $\vecu$, the optimal input cannot be found from \eqref{eq:Hu_general}. In many cases, this implies a ``bang-bang'' solution where the control switches between the upper and lower bounds, where the switching is determined by the sign of $\vecH_\vecu$.
On the other hand, in some cases when $\vecH_\vecu(t)$ can vanish for a finite time interval, the solution follows a singular arc and can be determined by the time derivatives:
\begin{equation}
 \frac{d^k}{dt^k}\vecH_\vecu=\vec{0}
 \label{Hu_time_der}
 \end{equation} 
After an even number of derivatives $k$, the input $\vecu$ appears and can be extracted \cite{bryson1975applied}. 
\\
\subsubsection*{Endpoints conditions}
Some OCPs may require initial-time and final-time conditions:
\begin{equation}
    \boldsymbol \Psi(t_0,\vecz_0,t_f,\vecz_f)=\mathbf{0}
\end{equation}
Where $\vecz_0\!=\!\vecz(t_0)$ and $\vecz_f\!=\!\vecz(t_f)$. In order to satisfy these conditions, the costate variables $\veclambda(t)$ must satisfy the transversality conditions:
\begin{equation}
\boldsymbol{\Lambda}=
    \begin{bmatrix} \veclambda(t_0)\\\veclambda(t_f) \end{bmatrix} = 
\left(\frac{d\boldsymbol{\Psi}}{d\textbf{Z}_{end}}\right)^T\textbf{L}+\left(\frac{dJ}{d\textbf{Z}_{end}}\right)^T
\label{eq:Tran_Con}
\end{equation}
Where $\textbf{Z}_{end}=[\vecz_0,\vecz_f]^T$ and $\textbf{L}\in\mathbb{R}^m$ is a vector of unknown scalar constants.
\\
\subsubsection*{Inequality constraints on state variables}

Consider an optimal control problem in which the state $\vecz(t)$ is also subject to an inequality constraint of the form:
\begin{equation}
w(\vecz(t))\leq 0
\label{eq:Bound_form}
\end{equation}
Following the \textit{direct adjoining approach} \cite{Speyer_Bryson_boundedstate,Hartl_survey_boundedstate}, a nonnegative Lagrange multiplier $\mu(t)\geq0$ associated with the constraint is introduced, and the augmented Hamiltonian is defined as:
\begin{equation}
\tilde H(\vecz,\vecu,\veclambda,\mu)=\veclambda^T \vecf+\mu w
\label{eq:Bounded_Hamiltonian}
\end{equation}
Such that an optimal solution satisfies:
\begin{equation}
\mu(t) w^*(\vecz(t))=0
\end{equation}
In time intervals where the constraint is inactive, i.e. $w(\vecz(t))<0$, the multiplier satisfies $\mu(t)=0$ and the optimal solution is found through $\vecH_\vecu=\boldsymbol{0}$ (or its time derivatives \eqref{Hu_time_der} in the case of a singular arc).
Suppose that the solution reaches the state constraint at an entry time $\tau_{1}$ such that $w(\vecz(\tau_1))=0$. For $t\geq\tau_{1}$, the optimal state evolves along the constraint surface $w(\vecz(t))=0$ while $\mu(t)>0$, where the control input $\vecu(t)$ is obtained from the time derivative of the constraint $\dot w=\nabla w^\top \vecf=0$ and the value of $\mu(t)$ arises from the OCP dynamics. The solution may later exit the constraint at a time $\tau_2>\tau_1$, such that $\mu(\tau_2)=0$. Then, for later times, $\mu(t>\tau_2)=0$, $w(t>\tau_2)<0$ and the solution reenters a constraint-free state trajectory. These alternating switches may occur several times along the optimal solution.
At such switching times, the costate variables may jump discontinuously. This jump is uniquely determined by the state constraint and satisfies $\veclambda(\tau^+)-\veclambda(\tau^-)=\mu(\tau) \frac{\partial w}{\partial \vecz}(\vecz(\tau))$ \cite{Speyer_Bryson_boundedstate}. In particular, any state variable $z_{i}$ that does not appear in the constraint (i.e. $\partial w/ \partial z_{i}=0$) has a corresponding costate component $\lambda_{i}(t)$ that changes continuously across a switching point.

\section{Maximal displacement gaits for a three-link kinematic swimmer}
\label{sec:Max_Dis}
In this section, we identify the gaits that maximize the net displacement of the robotic swimmer model described in \ref{sec:modified_model}. First, we examine simply-shaped displacement-suboptimal gaits and suggest the existence of multiple locally-optimal generally-shaped maximal displacement gaits. Second, we formulate the relevant OCP for a general three-link kinematic swimmer and solve it using PMP, similarly to \cite{wiezel2023geometric}. Then, we review the geometric analysis approach \cite{hatton2010optimizing}, utilized in \cite{wiezel2023geometric}, which allows us to visually assess the net displacement of a given input gait via a height map. Finally, we introduce the displacement-optimal gaits we discovered under different conditions, and use the geometric analysis results to further explain our findings.

\subsection{Motivation for multiple locally-optimal gaits maximizing net displacement}
\label{sec:Disp_Motivation}

\begin{figure}[ht]
\centering
\includegraphics[width=0.46\textwidth]{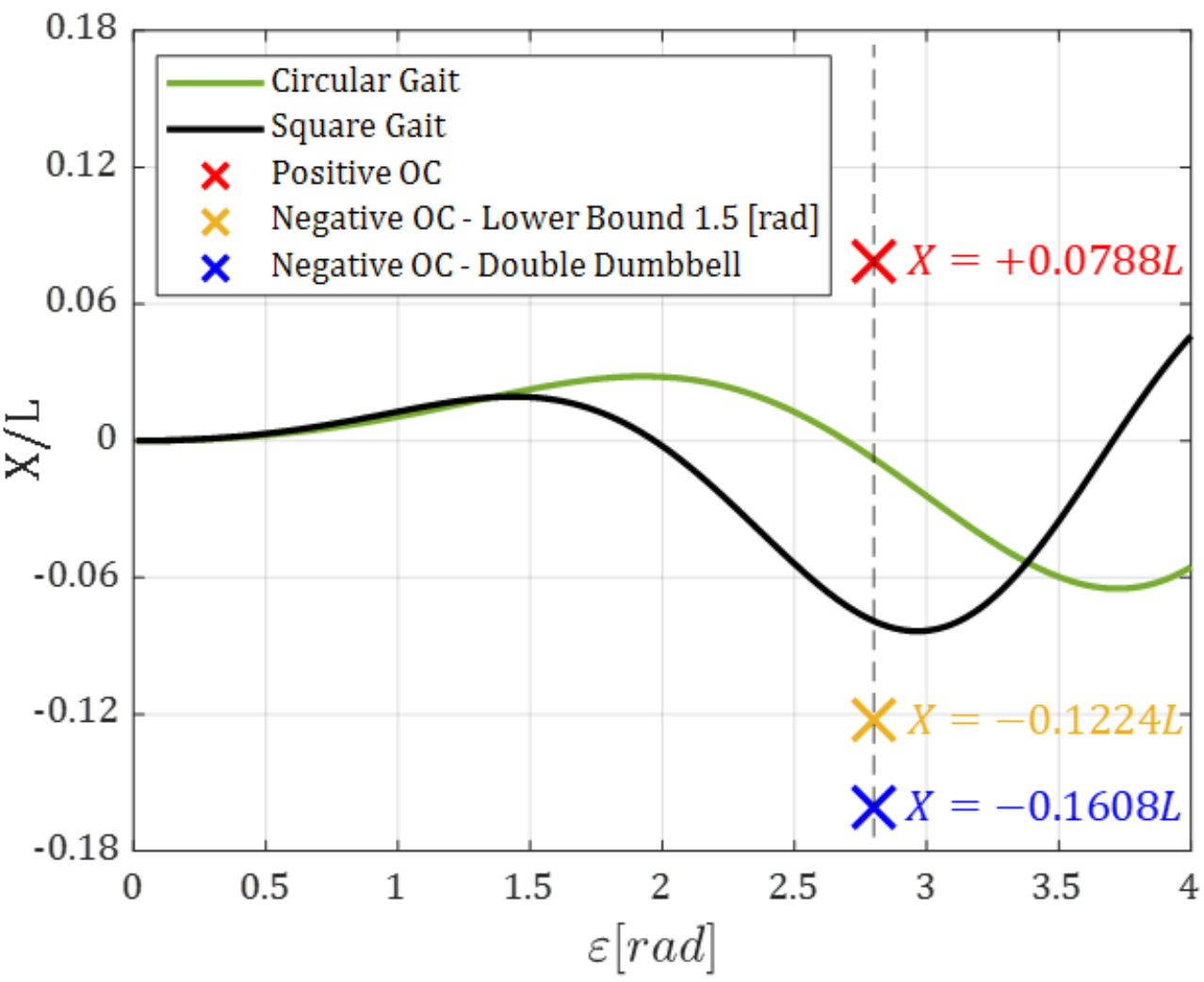}
\caption{Non-dimensional net displacement $X/L$ of the modified robotic macro-swimmer model, for varying amplitudes $\varepsilon$, where $L=l_0+l_1+l_2$ is the total links' length. Circular gaits in green and square-shaped gaits in black. Marked in x are the net displacement results of 3 locally-optimal generally-shaped gaits under an upper joint angle bound of $\phi_i\le2.8$ [rad], obtained by the OCP formulation detailed in \ref{sec:Disp_Res}.}
\label{Fig:Disp_Circs_and_Squares}
\end{figure}

As in the the work of Wiezel and Or \cite{wiezel2016optimization} for Purcell's swimmer, we examine the robotic swimmer's modified model behavior given circular and square-shaped input gaits with varying amplitudes, as shown in Fig. \ref{fig:Circ_and_Squ_Gait}. The plots in Fig. \ref{Fig:Disp_Circs_and_Squares} indicate the existence of two distinct displacement-optimal amplitudes for each shape, around 1.5-2 [rad] and 3-4 [rad]. For both shapes, the second optimum outperforms the first and also results in swimming in the opposite direction. As found in \cite{wiezel2023geometric} for Purcell's swimmer, this suggests that the robotic swimmer also has two distinct, locally-optimal, generally shaped gaits, associated with motion in opposite directions, i.e. positive and negative displacements. The existence of those gaits is the subject of research in the following subsections.   

%The graphs in Fig. \ref{fig:LHE_Purcell_Search_Square_and_Circle} show Purcell's swimmer Lighthill's efficiency  of circular and square-shaped gaits for varying amplitudes. For each shape, two distinct efficiency-optimal amplitudes are found, around 1.5 [rad] and 3 [rad]. In the circular gait, the first optimum performs better, whereas in the square gait, the second optimum outperforms the first and also surpasses the circular one.

%Displacement OCP formulation - start (To be IV.A)
\subsection{Formulation of the OCP for maximal displacement}
\label{sec:Disp_OCP_Formulation}
We formulate the relevant OCP similarly to \cite{wiezel2023geometric}. For a three-link kinematic swimmer, we define the state as $\vecz=[\phi_1,\phi_2,x,\theta]^T$, with the dynamics:

\begin{equation}
    \dot{\textbf{z}} = \begin{bmatrix} \dot\phi_{1}\\\dot\phi_{2}\\\dot x\\\dot\theta\\ \end{bmatrix} = \begin{bmatrix} u_1\\u_2\\g( \Phi,\theta)u_1+h(\Phi,\theta)u_2\\f(\Phi)u_1+q(\Phi)u_2 \end{bmatrix}
\label{eq:EOS_DISP}
\end{equation}

Where we denote the system's control input as $\vecu=[u_1, u_2]^T$ and the functions $f,g,h,q$ from (\ref{eq:EOS_DISP}) as elements of the matrix $\textbf{G}$ from (\ref{eq:full_connection}) such that $g(\Phi,\theta)=\textbf G_{11}, h(\Phi,\theta)=\textbf G_{12}, f(\Phi)=\textbf G_{31}, q(\Phi)=\textbf G_{32}$.
The cost function is now defined to maximize the net displacement of the swimmer in the $x$-direction over a period: 
\begin{equation}
J(t_f,\vecz_f)=x(t_f)
\end{equation}
Since the swimmer's body motion (\ref{eq:full_connection}) is time-invariant, its net displacement is independent of shape change rate and we can arbitrarily fix it as:
\begin{equation}
|\vecu(t)|=\sqrt{u_1^2+u_2^2}=1
\label{eq:Shape_Change_Rate}
\end{equation}
Similarly to previous analysis \cite{wiezel2016using,wiezel2023geometric}, we focus on gaits with symmtries about the two lines $\phi_1=\phi_2$ and $\phi_1=-\phi_2$, which yields net displacement in a single direction and no net rotation \cite{Gutman2015Symmetries}. Thus, we examine only one quarter of a gait, starting at $\phi_1=\phi_2$ and ending at $\phi_1=-\phi_2$. Hence, we define the following boundary conditions:
\begin{equation}
\boldsymbol \Psi\left(t_0,\vecz_0,t_f,\vecz_f\right)=\begin{bmatrix}
    \phi_1(t_0)-\phi_2(t_0) \\ x(t_0) \\ \theta(t_0) \\ \phi_1(t_f)+\phi_2(t_f)
\end{bmatrix}=\boldsymbol{0}
\label{eq:Trans_Con_Disp}
\end{equation}
The Hamiltonian is then defined from (\ref{eq:Hamiltonian_Def}) as:
\begin{equation}
H=\lambda_1u_1+\lambda_2u_2+\lambda_3(gu_1+hu_2)+\lambda_4(fu_1+qu_2)
\label{eq:Hamiltonian_Disp}
\end{equation}
and the costate dynamics are obtained from (\ref{eq:Costate_Dynamics_Def}) as:
\begin{equation}
\begin{array}{l}
    \dot \lambda_1=-\lambda_3(g_{\phi_1}u_1+h_{\phi_1}u_2)-\lambda_4(f_{\phi_1}u_1+q_{\phi_1}u_2) \\[12pt] \dot \lambda_2=-\lambda_3(g_{\phi_2}u_1+h_{\phi_2}u_2)-\lambda_4(f_{\phi_2}u_1+q_{\phi_2}u_2) \\[12pt] \dot \lambda_3=0 \rightarrow \lambda_3(t)=const \\[12pt] \dot \lambda_4 = -\lambda_3(g_{\theta}u_1+h_{\theta}u_2)
\end{array}
\label{eq:Costate_Disp}
\end{equation}
Where subscript letters denote partial derivatives, for example: 
$g_{\phi_1}=\frac{\partial g}{\partial \phi_1}$. From the transversality conditions (\ref{eq:Tran_Con}) we obtain the initial and final time conditions for the costate variables: 
\begin{equation}
\begin{array}{cc}            
\lambda_1(t_0)=-\lambda_2(t_0), \lambda_1(t_f)=\lambda_2(t_f)\\[5pt]
\lambda_3(t_f)=1, \lambda_4(t_f)=0
\end{array}
\label{eq:Trans_Con_Costate_Disp}
\end{equation}
The partial derivative of the Hamiltonian with respect to the control input is:
\begin{equation}
    H_\vecu=\begin{bmatrix}
        \lambda_1+g+\lambda_4f \\ \lambda_2+h+\lambda_4q
    \end{bmatrix}=\boldsymbol{0}
\label{eq:H_u_Disp}
\end{equation}
Since the Hamiltonian is linear with respect to the control input, the optimal control cannot be derived directly from $H_\vecu\!=\!\boldsymbol{0}$. However, a singular arc solution could be determined by an even number of time derivatives of $H_\vecu$. The first time derivative gives:
\begin{equation}
    \dot H_\vecu=F(\boldsymbol{\Phi},\theta,\lambda_3,\lambda_4)\begin{bmatrix}
        u^*_2\\-u^*_1
    \end{bmatrix}=\boldsymbol{0}
\label{eq:H_u_dot_Disp}
\end{equation}
Where:
\begin{equation}
\begin{array}{cc}            
F(\Phi,\theta,\lambda_3,\lambda_4)=F_1(\Phi,\theta)\lambda_3+F_2(\Phi,\theta)\lambda_4\\[5pt]
F_1(\Phi,\theta)=g_{\phi_2} + g_\theta q - h_{\phi_1} - h_\theta f \\[5pt]
F_2(\Phi,\theta)=f_{\phi_2} - q_{\phi_1}
\end{array}
\end{equation}
Therefore, the non-trivial optimal solution satisfies:
\begin{equation}
\lambda_4(\boldsymbol{\Phi},\theta,\lambda_3)=-\frac{F_1}{F_2}(\boldsymbol{\Phi},\theta)\lambda_3
\label{eq:lambda4_Disp}
\end{equation}
Next, from the second time derivative of $H_\vecu$, a relation between the optimal inputs arises: 
\begin{equation}
    \ddot H_\vecu=\dot F \begin{bmatrix}
        u^*_2\\-u^*_1
    \end{bmatrix}=\boldsymbol{0} \rightarrow \dot F = (Au^*_1+Bu^*_2)\lambda_3=0
\label{eq:H_u_ddot_Disp}
\end{equation}
Where:
\small
\begin{flalign}
A(\boldsymbol{\Phi},\theta)=(&q_{\phi_1\phi_1}-f_{\phi_2\phi_1})\frac{F_1}{F_2}+(2q_{\phi_1}-f_{\phi_2})g_{\theta}+\nonumber\\
(&g_{\phi_2\theta}+g_{\theta\theta}q-h_{\phi_1\theta}-h_{\theta\phi_1})f+g_{\phi_1\phi_2}+\nonumber\\
&g_{\theta\phi_1}q-h_{\theta\theta}f^2-h_{\theta}f_{\phi_1}-h_{\phi_1\phi_1}
\nonumber\\[6pt]
B(\boldsymbol{\Phi},\theta)=(&q_{\phi_1\phi_2}-f_{\phi_2\phi_2})\frac{F_1}{F_2}+(q_{\phi_1}-2f_{\phi_2})h_{\theta}+\nonumber\\
(&g_{\phi_2\theta}-h_{\theta\theta}f+g_{\theta\phi_2}-h_{\phi_1\theta})q+g_{\phi_2\phi_2}-\nonumber\\
&h_{\theta\phi_2}f+g_{\theta\theta}q^2+g_{\theta}q_{\phi_2}-h_{\phi_1\phi_2}
\end{flalign}
\normalsize
This relation essentially constitutes the direction of the local tangent to the optimal gait shape: 
\begin{equation}
u^*_2=-\frac{A(\boldsymbol{\Phi},\theta)}{B(\boldsymbol{\Phi},\theta)}u^*_1
\label{eq:Disp_Solution}
\end{equation}

The costate variables $\veclambda$ do not appear in the optimal control solution (\ref{eq:Disp_Solution})
and therefore their dynamics do not need to be solved. Consequently, we are left with a system of four ODEs, only for the state variables $(\phi_1,\phi_2,x,\theta)$. From
the boundary conditions (\ref{eq:Trans_Con_Disp}), only a single initial variable value remains unknown $\phi_1(t_0)=\phi_2(t_0)$. This results in a two-point boundary value problem (BVP) with a single initial guess variable $\phi_1(t_0)$ which, for the optimal solution, must satisfy one of the endpoint boundary conditions. From the costate dynamics (\ref{eq:Costate_Disp}), combined with the transversality conditions (\ref{eq:Trans_Con_Costate_Disp}), one obtains $\lambda_3=1$ $\forall t\in (t_0,t_f)$. Therefore, $\lambda_4$ can be calculated using (\ref{eq:lambda4_Disp}) and we use $\lambda_4(t_f)=0$ as the endpoint boundary condition of the BVP. This problem is solved using a simple scalar search, as in \cite{wiezel2023geometric}. 
\\[2pt]

\subsubsection*{Solution with bounded joint angles}
Assume a bound on the joints' angles $|\phi_i|\leq\phi_{max}$. Over a finite time interval with a nonzero control input, only the bound on one joint angle can be active, while the other joint angle is varying. Due to symmetry properties of the swimmer \cite{Gutman2015Symmetries}, we only consider one quadrant of the gait, where only one joint may reach the bound. Therefore, it is sufficient to consider only a single, scalar state bound of the form (\ref{eq:Bound_form}), which acts as an inequality state constraint. For our demonstration, we assume, without loss of generality, that the only bound is $w(\vecz)=\phi_2-\phi_{max}\leq0$. We modify the Hamiltonian from (\ref{eq:Hamiltonian_Disp}) according to (\ref{eq:Bounded_Hamiltonian}):
\begin{equation}
\tilde{H}=H+\mu(\phi_2-\phi_{max})
\label{eq:Disp_Modified_Hamiltonian}
\end{equation}
Where $\mu$ is an additional scalar multiplier. Consequently, the dynamics of the costate variable $\lambda_2$ is modified from (\ref{eq:Costate_Disp}) to:
\begin{equation}
\dot{\tilde\lambda}_2=\dot{\lambda}_2-\mu
\label{eq:Disp_Modified_lambda2dot}
\end{equation}
When the bound is inactive, $\phi_2(t)<\phi_{max}$, $\mu(t)=0$, and the optimal gait follows the unbounded optimal solution (\ref{eq:Disp_Solution}). When the bound is active $\phi_2(t)=\phi_{max}$, $\mu(t)>0$, and the optimal control input is obtained from the time derivative of the constraint $\dot w=u^*_2=0$, and (\ref{eq:Shape_Change_Rate}) gives $u^*_1=\pm1$. The entry time to the bound $\tau_1$ is simply the time when the unbounded optimal solution reaches the bound. The exit time from the bound $\tau_2$ is determined by a criterion arising from the optimal control formulation, accounting for the change in the costate dynamics (\ref{eq:Disp_Modified_lambda2dot}):
 \begin{equation}
\dot{\tilde{H}}_{\vecu}=F(\boldsymbol{\Phi},\theta,\lambda_3,\lambda_4)\begin{bmatrix}
     u^*_2\\-u^*_1
 \end{bmatrix}-\begin{bmatrix}
     0\\\mu
 \end{bmatrix}=\boldsymbol{0}
 \label{eq:Disp_Modified_Hu_dot}
 \end{equation}
%Displacement OCP formulation - end
The first element in the vector equation (\ref{eq:Disp_Modified_Hu_dot}) is trivially satisfied. Since the inequality constraint $w(\vecz)$ is independent of $\lambda_3$ and $\lambda_4$, the value of these costate variables is continuous, and from (\ref{eq:Costate_Disp}),(\ref{eq:Trans_Con_Costate_Disp}) one obtains $\lambda_3=1$ $\forall t\in (t_0,t_f)$ for the bounded solution as well. Now, by using the dynamics of $\lambda_4(t)$ from (\ref{eq:Costate_Disp}), one obtains the exit-time criterion from the second term of (\ref{eq:Disp_Modified_Hu_dot}):
\begin{equation}
F(\boldsymbol{\Phi},\theta,\lambda_3,\lambda_4)=\mu>0
\label{eq:Disp_Bound_Exit_Criterion}
\end{equation}
This means that when $F(\boldsymbol{\Phi},\theta,\lambda_3,\lambda_4)$=0, the optimal solution departs from the bound.

Below, in \ref{sec:Disp_Res}, we analyze cases involving two additional types of inequality state constraints. The first case imposes a lower joint angle bound $\phi_2\geq\phi_{min}$. The second case restricts the gait to remain within the quarter-gait sector $\phi_2\geq|\phi_1|$. In both cases, the analysis yields the same exit-time criterion function as presented above, differing only by the sign:

\begin{equation}
-F(\boldsymbol{\Phi},\theta,\lambda_3,\lambda_4)=\mu>0
\end{equation}

%Geometric analysis - start (To be IV.B)
\subsection{Geometric methods for finding maximal displacement gaits}
\label{sec:Geometric_Methods}
In order to get a better insight into the behavior of the kinematic models and understand the changes in the displacement-optimal gaits, we now review a geometric approach to the analysis of such swimmers, which has been introduced in \cite{hatton2010optimizing,hatton2011geometric} and utilized for OCP in \cite{wiezel2023geometric}.

The matrix $\vecG(\Phi,\theta)$ in (\ref{eq:full_connection}) can be decomposed \cite{Gutman2015Symmetries} into:
\begin{equation}
\vecG(\Phi,\theta)=\vecD(\theta)\vecA(\Phi)
\label{eq:Decompose_G}
\end{equation}
where $\vecD(\theta)$ is the rotation matrix:
\begin{equation}
\vecD(\theta)=
\begin{bmatrix}
\cos\theta & -\sin\theta & 0 \\
\sin\theta & \cos\theta & 0 \\
0 & 0 & 1
\end{bmatrix}
\label{eq:Rot_Mat}
\end{equation}

The net displacement of the swimmer over a period is equal to the line integral over the gait,
\begin{equation}
\Delta \vecq =
\ointctrclockwise_{\Phi}\vecD(\theta)\vecA\left(\Phi\right)
\label{eq:line_int}
\end{equation}
This line integral can be approximately converted to a surface integral in a manner similar to Stokes' theorem by evaluating the total Lie bracket $D\vecA$ of the system over the surface $\Phi_a$ as in \cite{ramasamy2016soap,ramasamy2019geometry}, giving the ``corrected Body Velocity Integral''

\begin{equation}
    \Delta \vecq \approx \iint_{\Phi_a}D\vecA = \iint_{\Phi_a}d\vecA-[\vecA_1,\vecA_2]
    \label{eq:surfint}
\end{equation}

Here, $d\vecA$ is the exterior derivative of the local connection (the generalized row-wise curl), and $[\vecA_1,\vecA_2]$ is a local Lie bracket term that corrects for noncommutativity as the swimmer translates and rotates through space. 

Note here that although the local Lie bracket term compensates for first-order noncommutativity effects, it is not a perfect correction, and the residual error grows proportionally with the amount of intermediate rotation during the gait. 

The choice of body-dependent coordinates affects the magnitude of this residual error. The frame fixed to the central link rotates at the angle $\theta(t)$ that becomes large for large-amplitude gaits, and so residual error will grow proportionally quickly. 
However, other shape-dependent frames can be chosen such that the rotation of the frame is minimal. For the swimmers we consider here, good frames are approximately at the center of mass and aligned with the mean orientation of the links: a general algorithm for finding good body frames is presented in \cite{hatton2011geometric}. 

\subsubsection*{Height Functions and Displacement-Optimal Gaits}
Using the ``minimum-perturbation coordinates'', gives a good approximation of the swimmer's net displacement over a gait. By plotting the x-integrand in \eqref{eq:surfint} as a height function $H_x\left(\Phi\right) = D\vecA^x(\Phi)$, one can identify sign-definite areas of the shape plane. A positive area that is encircled by the gait (in a counter-clockwise direction) will result in positive net displacement of the swimmer in the $x$ direction. 
In order to maximize the displacement of the swimmer, the gait should enclose a region of the height function that is as sign-definite as possible. Obviously, this will be accomplished by following the \textit{zero-level curve} that separates between positive and negative regions. For the symmetric three-link swimmer with slender links, the zero-level curve found in \cite{wiezel2023geometric} is nearly identical to the displacement-optimal gait found numerically in \cite{tam2007optimal} and reproduced using PMP in \cite{wiezel2016using}. 
%Geometric analysis - end

\subsection{Results for maximal net displacement of the modified robotic model}
\label{sec:Disp_Res}
We now present the results of the OCP for maximal net displacement of the robotic swimmer's modified model, as proposed in \ref{sec:modified_model}.

\begin{figure}[t]
\centering
\begin{subfigure}[c]{0.235\textwidth}
\includegraphics[width=\textwidth]{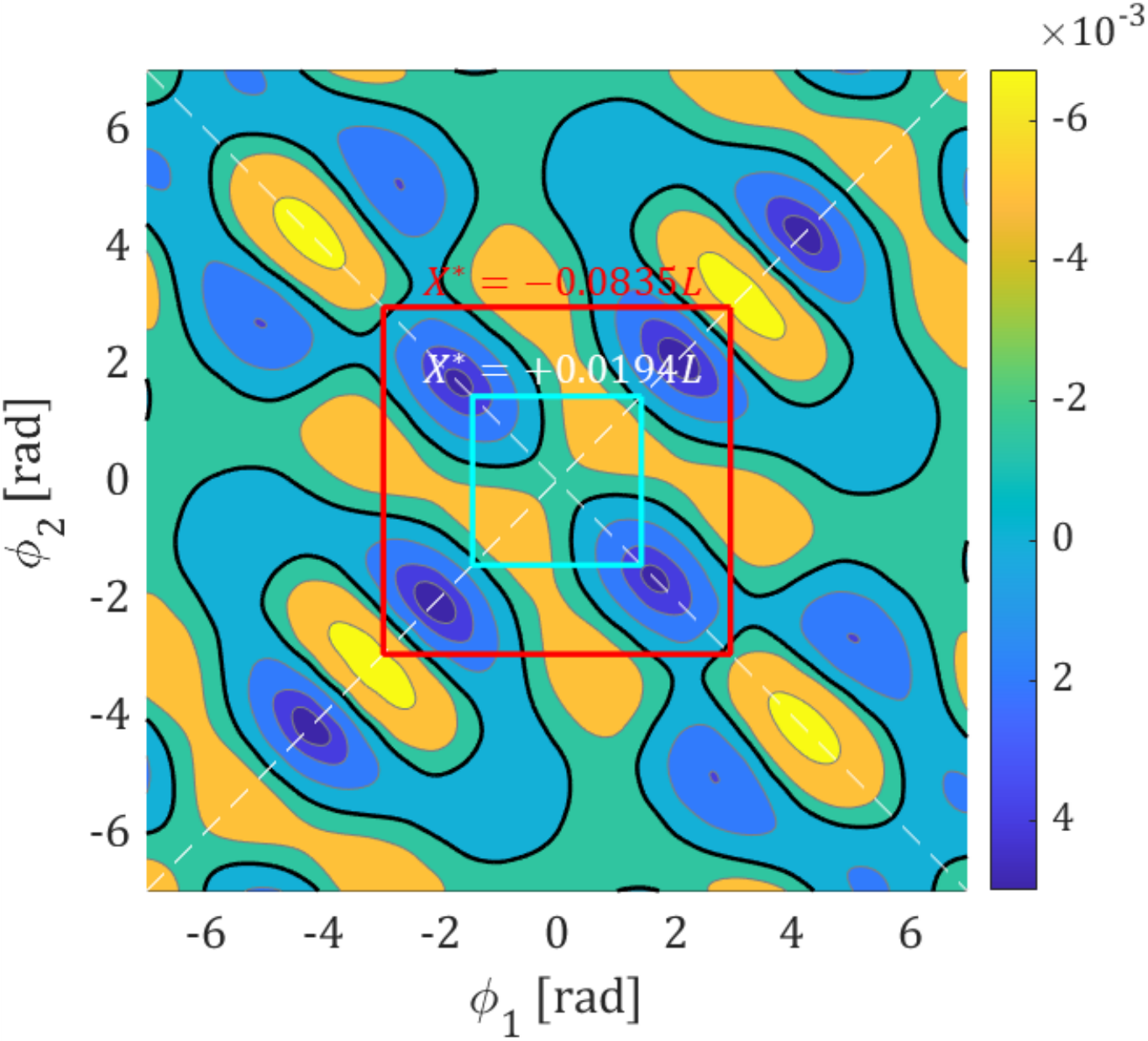}
\caption{}
\label{fig:Dis_HM}
\end{subfigure}
\begin{subfigure}[c]{0.235\textwidth}
\includegraphics[width=\textwidth]{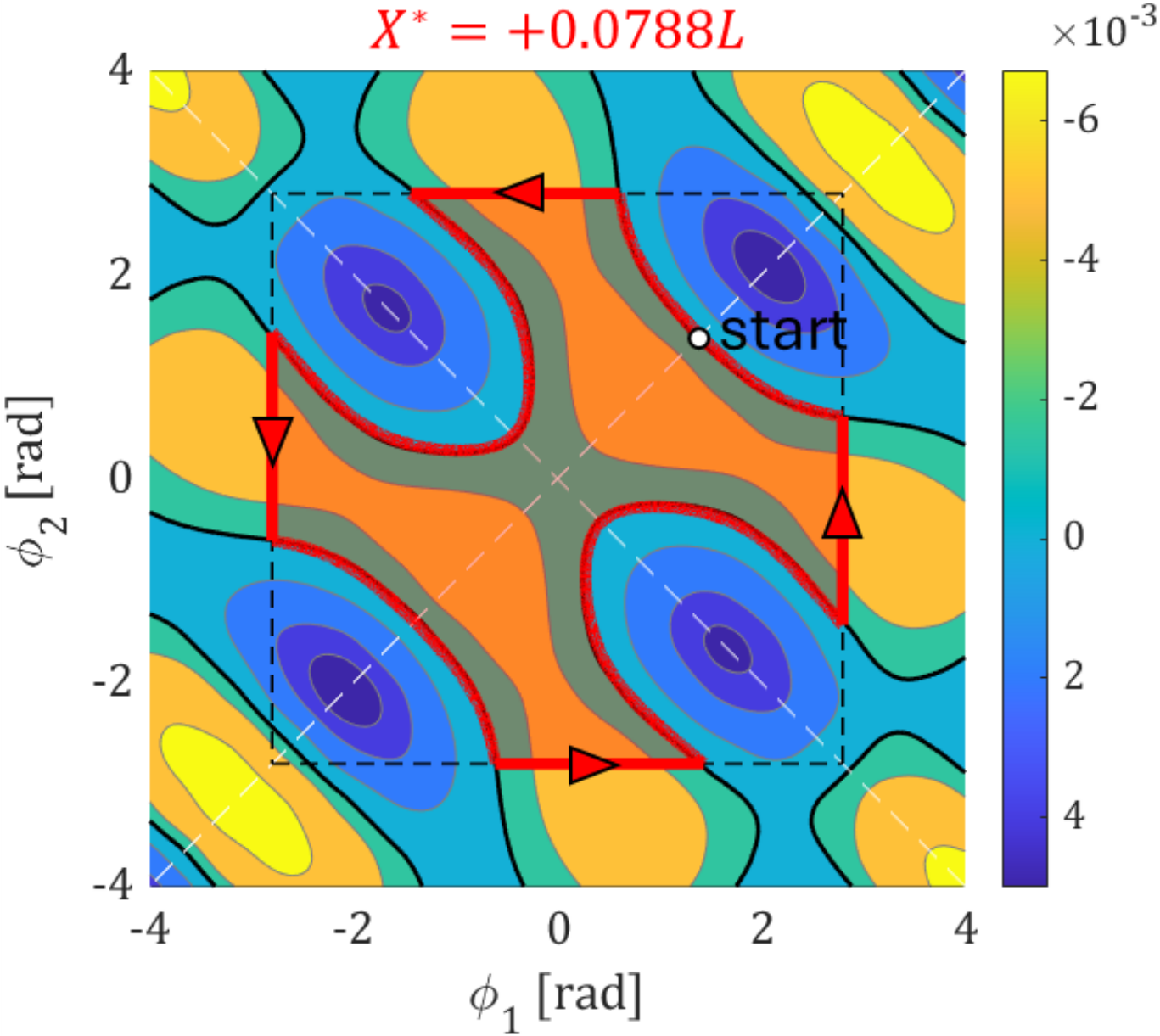}
\caption{}
\label{fig:Dis_Res_1}
\end{subfigure}
 \begin{subfigure}[c]{0.235\textwidth}
\includegraphics[width=\textwidth]{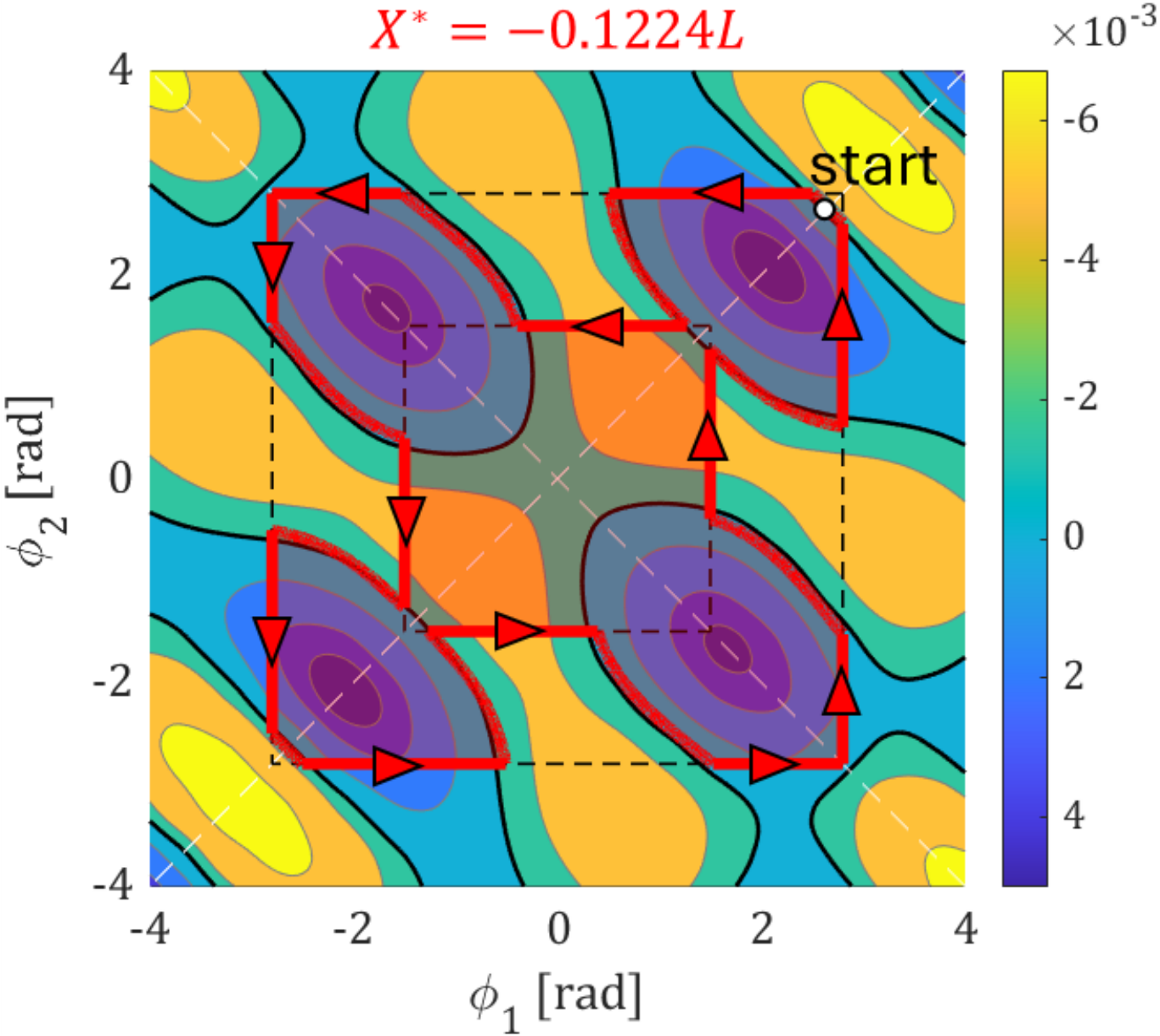}
\caption{}
\label{fig:Dis_Res_2}
\end{subfigure}
\begin{subfigure}[c]{0.235\textwidth}
\includegraphics[width=\textwidth]{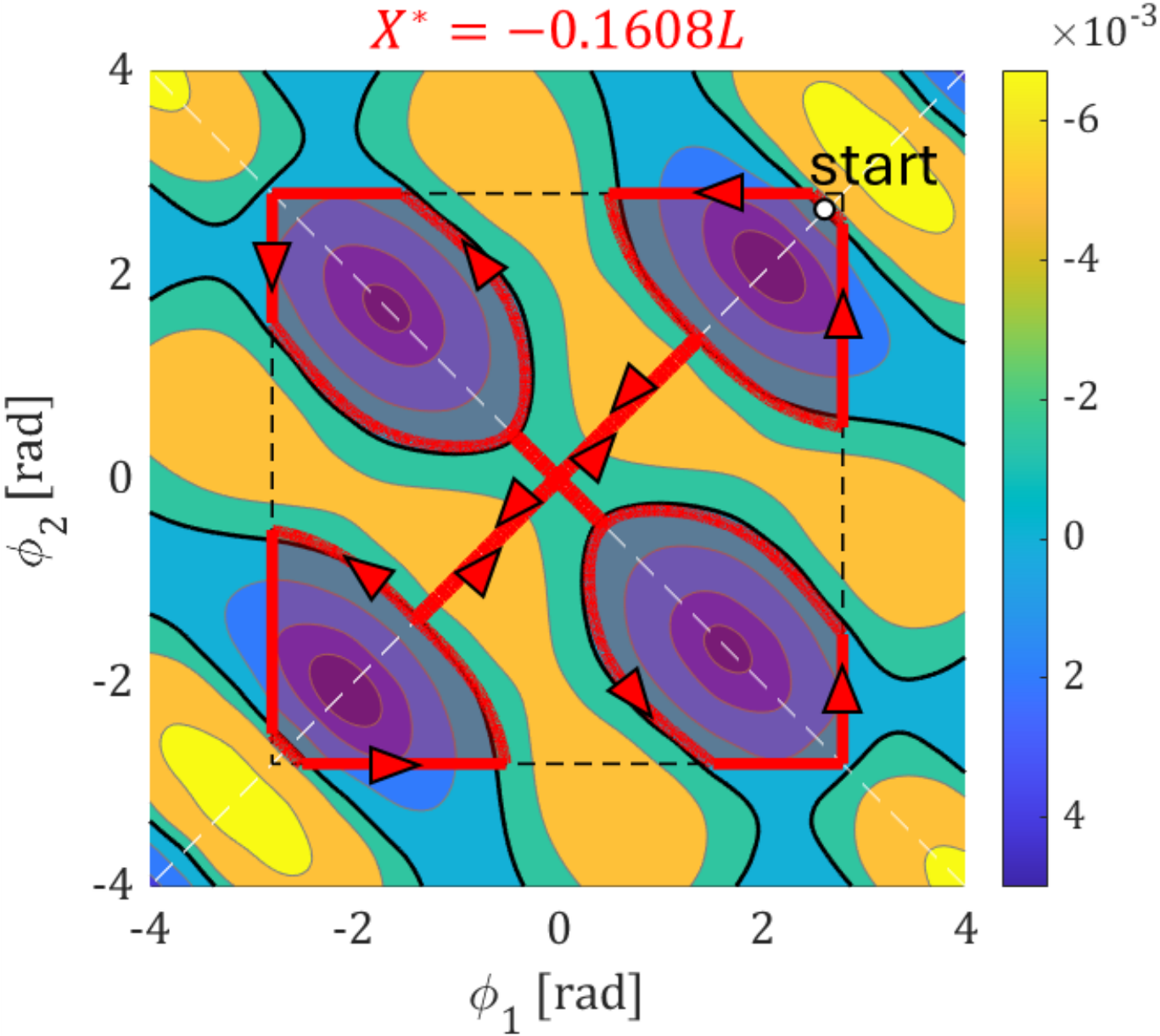}
\caption{}
\label{fig:Dis_Res_3}
\end{subfigure}
\caption{(a) Height map of the net displacement of the robotic swimmer's modified model and its zero-level curves (black curves), calculated using \cite{HattonSoftware}. Overlaid are the first and second sub-optimal square gaits (white and red lines respectively). (b) Locally-optimal gait with positive net displacement under an upper joint bound of $|\phi_i|\le2.8$ [rad]. \\(c) Locally-optimal gait with negative net displacement under lower and upper joint bounds of $1.5\le|\phi_i|\le2.8$ [rad]. \\(d) Locally-optimal gait with negative net displacement under an upper joint bound of $|\phi_i|\le2.8$ [rad] and inequality state constraints of the form $|\phi_i|\ge|\phi_j|$, which results in a double-dumbbell loop.}
\label{fig:Results_Disp}
\end{figure}

First, we found that without incorporating bounds, PMP fails to find displacement-optimal gaits, similarly to Purcell's swimmer large-amplitude displacement-optimal gait \cite{wiezel2023geometric}. This can be explained by examining the system's corresponding height function, shown in Fig. \ref{fig:Dis_HM}, acquired using \cite{HattonSoftware}, as detailed in \ref{sec:Geometric_Methods}. It is evident that no continuous curve on the $\phi_1\!-\!\phi_2$ plane can simultaneously follow the height map's zero-level curves and satisfy the endpoint conditions $\phi_1(t_0)\!=\!\phi_2(t_0)$ and $\phi_1(t_f)=$$-\phi_2(t_f)$. Therefore, the unbounded OCP does not have a unique solution, and introducing bounds is necessary in order to obtain locally-optimal solutions.

This height map also explains the displacement-optimal square-shaped gaits in Fig. \ref{Fig:Disp_Circs_and_Squares}. The first optimum mainly encircles positive net displacement regions. Increasing the square's amplitude gradually captures more negative regions, causing the net displacement to decrease, change sign, and reach a second, negative optimum.

Applying an upper bound on the OCP allows us to find a positive net displacement locally-optimal gait, as demonstrated in Fig. \ref{fig:Dis_Res_1} for $|\phi_i|\le2.8$ [rad]. This gait closely follows the zero-level curves, and is only trimmed when it reaches the bound. The segment that follows the bound acts as a “bridge” between disconnected zero-level curves. Note that this gait fully encloses the positive area of the height map bounded by the upper joint bound while avoiding any negative areas, thus maximizing the positive net displacement of the swimmer under such bound. 

We now present the analysis of a case in which both lower and upper bounds are imposed on the joint angles. In this case, the resulting displacement-optimal gait may contain multiple transitions between singular-arc segments and segments evolving along the upper or lower bounds of the joint angles, allowing us to enclose an increased amount of sign-definitive area of the height map. We demonstrate this behavior for the bounds $1.5\le|\phi_i|\le2.8$ [rad], with the resulting displacement-optimal gait shown in Fig. \ref{fig:Dis_Res_2}. This gait results in negative net displacement that outperforms its positive counterpart from Fig. \ref{fig:Dis_Res_1} by a 55\% (in absolute value). 

We now consider another case, with an upper bound $\phi_2\le \phi_{max}$ and $\phi_2\ge|\phi_1|$. The latter inequalities bound the angles within a quarter-gait sector, while allowing inclusion of segments along its boundary lines, the two diagonals $\phi_2=\pm\phi_1$. The resulting locally-optimal gait for $\phi_{max}=2.8$ [rad] is shown in Fig. \ref{fig:Dis_Res_3}. It can be seen that this is a double-dumbbell loop, similar to the one considered in \cite{avron2008comment}, that fully encircles only the negative regions in the height map. This gait outperforms its positive and negative counterparts from Fig. \ref{fig:Dis_Res_1} and \ref{fig:Dis_Res_2} by 104\% and 31\% increase in absolute net displacement, respectively. 

Finally, Fig. \ref{Fig:Disp_Circs_and_Squares} in \ref{sec:Disp_Motivation} above, compares the performance of these gaits to those of circular and square shaped gaits. 

\section{Maximal Lighthill's efficiency gait}

In this section, we identify the gaits that maximize Lighthill's energy efficiency for both Purcell's swimmer model and the robotic swimmer model. First, we examine simply-shaped efficiency-suboptimal gaits and suggest, as in the case of displacement optimization, existence of multiple locally-optimal generally-shaped gaits maximizing Lighthill's energy efficiency. Next, we formulate the relevant OCP for a general three-link kinematic swimmer and its numerical solution scheme using PMP. Finally, we introduce the new efficiency-optimal gaits we discovered for both swimmers using this method. 

\subsection{Motivation for multiple locally-optimal gaits maximizing Lighthill's energy efficiency}
\label{sec:LHE_Motivation}
\begin{figure}[t]
\centering
\begin{subfigure}[c]{0.235\textwidth}
\includegraphics[width=\textwidth]{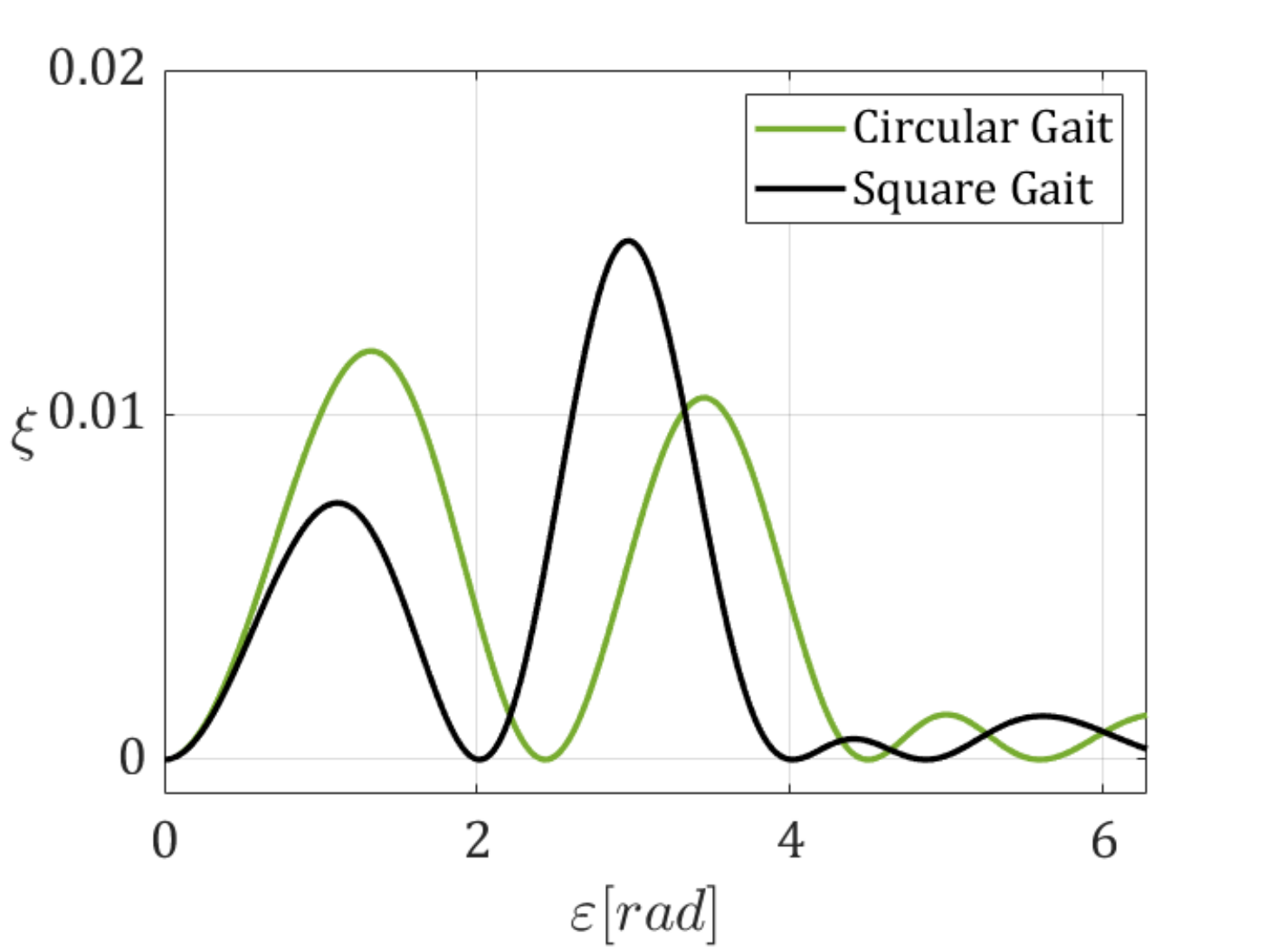}
\caption{}
\label{fig:LHE_Purcell_Search_Square_and_Circle}
\end{subfigure}
\begin{subfigure}[c]{0.235\textwidth}
\includegraphics[width=\textwidth]{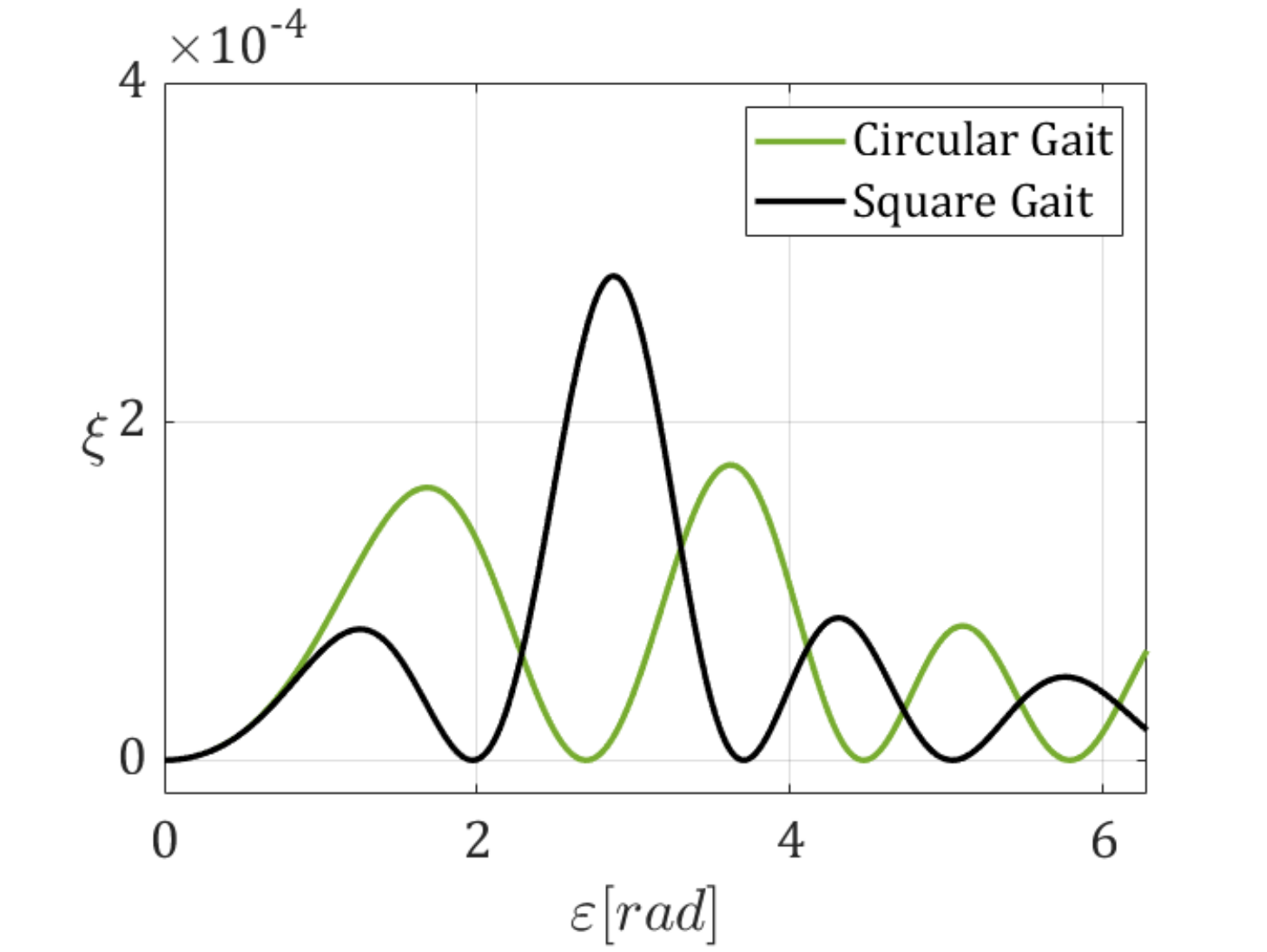}
\caption{}
\label{fig:LHE_Buoy_Search_Square_and_Circle}
\end{subfigure}
\caption{Lighthill’s efficiency $\xi$ for varying amplitudes $\varepsilon$. Circular gaits in green and square-shaped gaits in black. (a) Purcell's swimmer. (b) The modified robotic macro-swimmer model.}
 \label{fig:LHE_Circs_and_Squares}
\vspace{-2mm}
\end{figure}

Similarly to the net displacement analysis, we first examine Lighthill's energy efficiency for both Purcell's swimmer and the robotic swimmer models, under circular and square-shaped input gaits with varying amplitudes, depicted in Fig. \ref{fig:LHE_Circs_and_Squares}. We notice that the non-dimensional Lighthill's energy efficiency of the robotic swimmer model is two orders of magnitude smaller than that of Purcell's slender-links swimmer. This is due to the lower net displacement of the robotic swimmer, as well as its increased drag structure, resulting in higher energy dissipation. Following the net displacement results for both swimmers, two distinct efficiency-optimal simply-shaped gaits are found: a smaller circular gait around an amplitude of 1-2 [rad] and a larger square gait around an amplitude of 3-4 [rad]. For both swimmers, the second optimum outperforms the first and is associated with negative net displacement. This suggests that both swimmers may also have two distinct, locally-optimal, generally shaped gaits. The existence of those gaits is the subject of research in the following subsections.

%LHE OCP formulation - start (To be V.A)
\subsection{Formulation of OCP for maximizing Lighthill's efficiency}
\label{sec:eff_OCP}
In order to formulate the OCP with Lighthill’s efficiency $\xi$ as the objective function, we expand the state vector in (\ref{eq:EOS_DISP}) to include the swimmer's energy expenditure $E(t)$. The system's state will now be $\textbf{z}=[\phi_1,\phi_2, x, \theta, E]^T$ and the state dynamics are now:
\begin{align}
    \dot{\textbf{z}} = \begin{bmatrix} \dot\phi_{1}\\\dot\phi_{2}\\\dot x\\\dot\theta\\\dot E \end{bmatrix} = \begin{bmatrix} u_1\\u_2\\g( \Phi,\theta)u_1+h(\Phi,\theta)u_2\\f(\Phi)u_1+q(\Phi)u_2\\a(\Phi)u_1^2+b(\Phi)u_1u_2+c(\Phi)u_2^2 \end{bmatrix}
\label{eq:LHE_EOS}
\end{align}

where $a(\Phi)=\textbf W_{11}$, $\,b(\Phi)=2\textbf W_{12}$ and $c(\Phi)=\textbf W_{22}$
are the elements of the matrix $\textbf W(\Phi)$ from (\ref{eq:power}). The cost function is now defined proportionally to Lighthill’s
energy efficiency $\xi$ in (\ref{eq:LHE_DEF}) as:
\begin{equation}
    J(t_f,\vecz_f)=\frac{x^2(t_f)}{E_ft_f}
    \label{eq:LH_CF}
\end{equation}
Similarly to the maximal displacement analysis \ref{sec:Disp_OCP_Formulation}, we focus on symmetric gaits and examine
only one quarter of the gait. Hence, We define the suitable endpoints state conditions:
\begin{align}
    \boldsymbol \Psi\left(t_0,\vecz_0,t_f,\vecz_f\right) = \begin{bmatrix} \phi_1(0)-\phi_2(0)\\x(0)\\\theta(0)\\E(0)\\\phi_1(t_f)+\phi_2(t_f)\\t_f-1 \end{bmatrix} = \boldsymbol{0}
\label{eq:LHE_Endpoints_Con}
\end{align}
Where the final time is chosen arbitrarily as $t_f=1$ in order to fully define the OCP. The Hamiltonian is then defined from (\ref{eq:Hamiltonian_Def}) and (\ref{eq:LHE_EOS}) as:
\begin{equation}
\begin{array}{c}
H=\lambda_1u_1+\lambda_2u_2+\lambda_3(gu_1+hu_2)+\\[6pt]  \lambda_4(fu_1+qu_2)+ \lambda_5(au_1^2+bu_1u_2+cu_2^2)
\end{array}
\end{equation}
and the costate dynamics are obtained from (\ref{eq:Costate_Dynamics_Def}) as:
\begin{equation}
\begin{array}{l}
    \dot\lambda_1=-\lambda_3(g_{\phi_1}u_1+h_{\phi_1}u_2)-\lambda_4(f_{\phi_1}u_1+q_{\phi_1}u_2)\\[4pt] \ \qquad-\lambda_5(a_{\phi_1}u_1^2+b_{\phi_1}u_1u_2+c_{\phi_1}u_2^2)\\[8pt]
    \dot\lambda_2=-\lambda_3(g_{\phi_2}u_1+h_{\phi_2}u_2)-\lambda_4(f_{\phi_2}u_1+q_{\phi_2}u_2)\\[4pt] \ \qquad-\lambda_5(a_{\phi_2}u_1^2+b_{\phi_2}u_1u_2+c_{\phi_2}u_2^2)\\[8pt]
    \dot\lambda_3=0\rightarrow \lambda_3(t)=const\\[8pt]
    \dot\lambda_4=-\lambda_3(g_{\theta}u_1+h_{\theta}u_2)\\[8pt]
     \dot\lambda_5=0\rightarrow \lambda_5(t)=const
\end{array}
\label{eq:LHE_Costate_Dynamics}
\end{equation}
In contrast to the maximal displacement analysis, here the Hamiltonian depends nonlinearly on the control input. Thus, we can derive the efficiency-optimal control input directly from PMP. The partial derivative $H_{\boldsymbol{u}}$ is:
\begin{equation}
    H_{\boldsymbol{u}} = \begin{bmatrix} \lambda_1+\lambda_3 g+\lambda_4f+\lambda_5(2au_1+bu_2)\\\lambda_2+\lambda_3 h+\lambda_4q+\lambda_5(bu_1+2cu_2) \end{bmatrix}
\label{eq:Hu_LHE}
\end{equation}
And the optimal control is obtained from solving $H_{\boldsymbol{u}}\!=\!\boldsymbol{0}$:
\begin{equation}
\textbf{u}^*=\gamma^{-1}\begin{bmatrix}
b\alpha-2c\beta\\b\beta-2a\alpha   
\end{bmatrix}
\label{eq:LHE_u*}
\end{equation}
where:
\begin{equation}
\begin{array}{cc}            
    \alpha=\lambda_2+\lambda_3h+\lambda_4q \\[5pt]
    \beta=\lambda_1+\lambda_3g+\lambda_4f \\[5pt]
    \gamma=\lambda_5(4ac-b^2)
\end{array}
\end{equation}
From the transversality conditions \eqref{eq:Tran_Con} one obtains the initial and final time conditions for the costate variables:
\begin{equation}
\begin{array}{cc}
    \lambda_1(t_0)=-\lambda_2(t_0), \lambda_1(t_f)=\lambda_2(t_f), \\[8pt]\lambda_3(t_f)=-\frac{2x(t_f)}{E(t_f)},
    \lambda_4(t_f)=0, \lambda_5=-\frac{x^2(t_f)}{E^2(t_f)}
    \label{eq:Trans_Con_LHE}
    \end{array}
\end{equation}
Then, combined with the costate dynamics (\ref{eq:LHE_Costate_Dynamics}) one obtains:
\begin{equation}
    \lambda_3(t)=2\frac{x_f}{E_f}, \lambda_5(t)=-\frac{\lambda_3^2}{4}
\label{eq:LHE_lambda_3_and_5}
\end{equation}
From this formulation, two main properties of an optimal Lighthill's efficiency gait arise:
\begin{enumerate}[I.]
    \itemsep0.1em
    \item \label{itm:constant-power}
    Constant power expenditure – As Becker et al. showed in \cite{becker2003self}, a gait with maximal Lighthill's efficiency must always maintain a constant power expenditure $p(t)$. In Appendix \ref{sec:Appendix_A} we use the optimal control formulation to prove that this property must apply for any general low-$Re$ three-link swimmer. 
    \item \label{itm:smoothness}
    Smoothness at the endpoints – since we use a symmetric gait and portray only a quarter of it, the gait might not be smooth at the endpoints (mirroring the gait about the axes of symmetry may result in a non-smooth curve). In Appendix \ref{sec:Appendix_B} we use the optimal control formulation to prove that a gait with maximal Lighthill's efficiency must be smooth at its ends, i.e. $u_1(t_0)\!=\!-u_2(t_0)\!<\!0$ and $u_1(t_f)\!=\!u_2(t_f)\!<\!0$, for a symmetric $(l_1\!=\!l_2)$ low-$Re$ three-link swimmer. 
\end{enumerate}
We use these properties in \ref{sec:LHE_BVP_Sol} to obtain optimal solutions for maximal Lighthill's efficiency.
%LHE OCP formulation - end

%LHE OCP BVP solution - start (To be V.B)
\subsection{Lighthill's efficiency OCP solution}
\label{sec:LHE_BVP_Sol}
Equations (\ref{eq:LHE_EOS}),(\ref{eq:LHE_Endpoints_Con}),(\ref{eq:LHE_Costate_Dynamics}),(\ref{eq:LHE_u*})-(\ref{eq:LHE_lambda_3_and_5}) give a BVP with four initial time $t_0$ guess values:
\begin{equation}
\phi_1(t_0)\!=\!\phi_2(t_0),\lambda_1(t_0),\lambda_3,\lambda_4(t_0)
\label{eq:In_Guess_LHE}
\end{equation}
This shooting problem is highly sensitive to the initial guess values and choosing them randomly is insufficient to guarantee that a nonlinear solver will converge to the correct solution. Therefore, we wish to choose our initial guess values as the initial values of a ``good enough'' gait – i.e. the optimum of a parametrized gait of a known shape, such as truncated Fourier series, as shown in Appendix \ref{sec:Appendix_C}. However, not all four initial time guess values in (\ref{eq:In_Guess_LHE}) can be directly obtained from a known gait, since three of them are costate variables rather than physical state variables. A resolution of this problem is obtained by guessing final time values at $t=t_f$ and integrating the system backward in time down to $t\!=\!t_0$. This is possible since all four final time guess variables could be physical state variables, which are acquired from the ``good enough'' gait:
\begin{equation}
 \phi_1(t_f)\!=\!-\phi_2(t_f),x(t_f),\theta(t_f),E(t_f)
\label{eq:Fin_Guess_LHE}
\end{equation}
This is achieved by using the properties of constant power expenditure $\dot p\!=\!E(t_f)/t_f$ and smoothness at the endpoints $u_1(t_f)\!=\!u_2(t_f)\!<\!0$ in order to acquire the final time costate variables values. Substituting into (\ref{eq:power}) one obtains:
 \begin{equation}
 \label{eq:u1tf_eq_u2tf}
 \begin{array}{cc}
      E(t_f)=(a+b+c)u_1^2|_{t=t_f}\\[8pt]
      \rightarrow u_1(t_f)=u_2(t_f)=\left. -\sqrt{\frac{E}{a+b+c}}\right|_{t=t_f}
      \end{array}
 \end{equation}
 Where we choose the negative sign of the square root in order to continue in the correct direction pointing to the next quarter of the gait.
 Substituting (\ref{eq:u1tf_eq_u2tf}) into (\ref{eq:Hu_LHE}) one obtains:
 \begin{equation}
 \label{eq:lambda1_2_LHE}
      \lambda_1(t_f)=\lambda_2(t_f)=-\lambda_3g-\lambda_5(2a+b)u_1|_{t=t_f}
 \end{equation}
 From (\ref{eq:lambda1_2_LHE}) and the transversality conditions (\ref{eq:Trans_Con_LHE}), the final time values of all costate variables are now known, and all final time guess values are physical state variables. For an efficiency-optimal gait, these four values, integrated backwards in time, must satisfy our four initial time conditions:
 \begin{equation}
 \phi_1(t_0)\!=\!\phi_2(t_0),x(t_0)\!=\!0,\theta(t_0)\!=\!0,\lambda_   1(t_0)\!=\!\lambda_2(t_0)
\label{eq:Ini_Cond_LHE}
\end{equation}
This shooting problem is then solved using \textit{Matlab}'s nonlinear solver \textit{fsolve}. We now demonstrate how, using the shooting method described above, we compute optimal gaits for Lighthill's efficiency. First, we present the results for the original Purcell swimmer, followed by the modified robotic swimmer model.
%LHE OCP BVP solution - end

%Purcell Swimmer Optimal LHE results - start (To be V.C)
\subsection{Lighthill's efficiency OCP results for Purcell's swimmer model}
\label{sec:Purcell_LHE_Results}
We now present the results of our optimal control analysis on efficiency-optimal gaits for Purcell's swimmer. The analysis confirms the existence of two distinct generally-shaped, locally-optimal gaits maximizing Lighthill's energy efficiency, as suggested in Fig. \ref{fig:LHE_Purcell_Search_Square_and_Circle}. 

The first generally-shaped efficiency-optimal gait we found, as shown in the red curve in Fig. \ref{fig:LHE_Optimal_Gaits_Purcell}, was obtained from an initial guess of the efficiency-optimal circular gait and closley resembles it. This gait is identical to Tam and Hosoi's \cite{tam2007optimal} efficiency-optimal gait, which was found numerically using truncated Fourier series, and thus we analytically confirmed the local optimality of this gait via PMP analysis.   

Furthermore, our approach identifies a second, large-amplitude, locally-optimal gait, shown in the blue curve in Fig. \ref{fig:LHE_Optimal_Gaits_Purcell}. This negative net displacement gait has a 46\% higher Lighthill's efficiency than the first generally-shaped efficiency-optimal gait. Interestingly, this gait resembles the locally-optimal square-shaped gait, though the PMP gait improves Lighthill energy efficiency by 20\% over this square gait. Importantly, a significant challenge was finding this efficiency-optimal gait using the shooting method, which is numerically sensitive and requires an initial guess based on the efficiency-optimal $3^{rd}$-order Fourier series gait (see Appendix \ref{sec:Appendix_C}).

%V.C - Purcell Swimmer Optimal LHE results - end
%\setlength{\belowcaptionskip}{-1pt}

%V.D - Robotic Swimmer Optimal LHE results - start
\subsection{Lighthill's efficiency OCP results for the robotic swimmer's modified model}
\label{sec:Robotic_LHE_Results} 
We now present the results of our optimal control analysis on efficiency-optimal gaits for the robotic swimmer. Similarly to Purcell's swimmer, the optimal control analysis results of the robotic swimmer's modified model confirms the existence of two distinct generally-shaped, locally-optimal gaits maximizing Lighthill's energy efficiency, as suggested in Fig. \ref{fig:LHE_Buoy_Search_Square_and_Circle}.

\begin{figure}[t]
\centering
\begin{subfigure}[c]{0.235\textwidth}
\includegraphics[width=\textwidth]{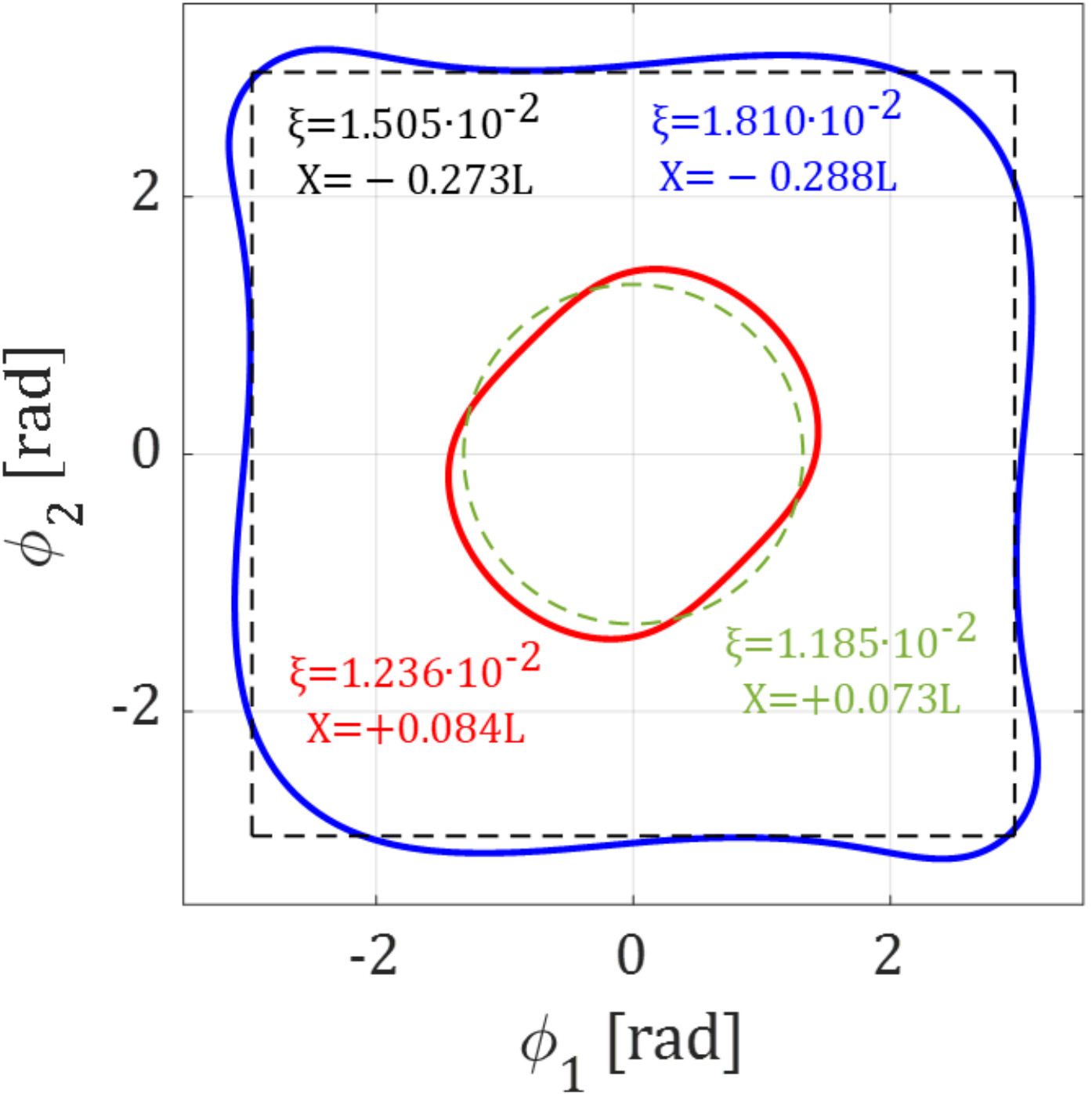}
\caption{}
\label{fig:LHE_Optimal_Gaits_Purcell}
\end{subfigure}
\begin{subfigure}[c]{0.235\textwidth}
\includegraphics[width=\textwidth]{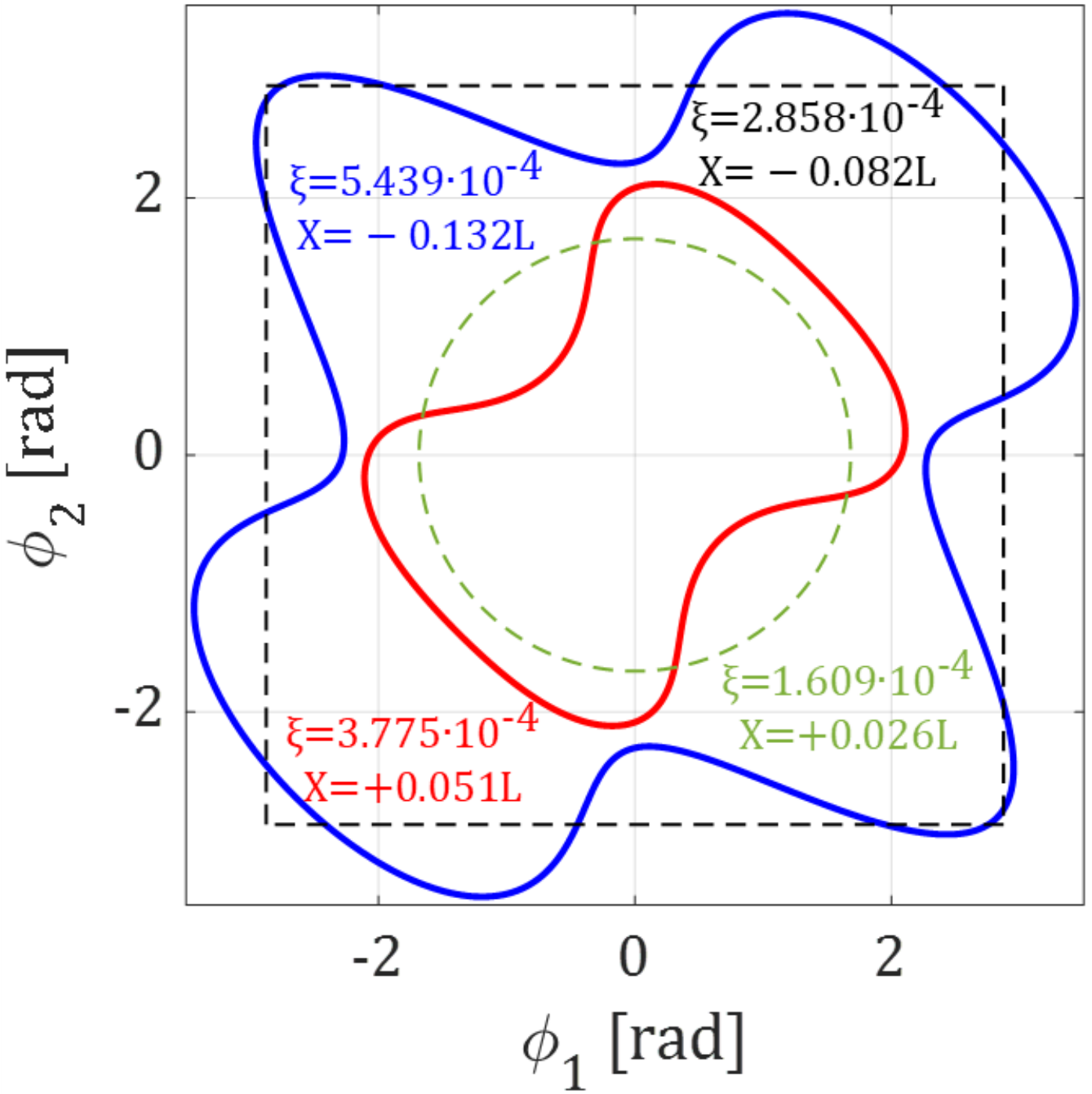}
\caption{}
\label{fig:LHE_Optimal_Gaits_Robot}
\end{subfigure}
\caption{Efficiency-optimal gaits, as produced by optimal control analysis, and their simply shaped suboptimal counterparts in dashed lines. (a) Purcell's swimmer. (b) The modified robotic swimmer's model.}
 \label{fig:Results_LHE}
 \vspace{-2mm}
\end{figure}

Our optimal control analysis manages to produce the first, low-amplitude generally shaped efficiency-optimal gait, as shown in the red curve in Fig. \ref{fig:LHE_Optimal_Gaits_Robot}. The shape of this gait differs significantly from the Purcell's swimmer counterpart, and exhibits an area of a noticeable cavity. This characteristic causes our optimal control method to converge only from an initial guess of the efficiency-optimal $2^{nd}$-order Fourier series.

As indicated by the results in Fig. \ref{fig:LHE_Buoy_Search_Square_and_Circle} for circular and square-shaped gaits, we expect another locally-optimal gait at a larger amplitude. We were able to obtain such gait, shown in the blue curve in Fig. \ref{fig:LHE_Optimal_Gaits_Robot}. However, it was difficult to converge to this gait using initial guess of optimal Fourier coefficients, due to high numerical sensitivity of the BVP. In order to overcome this, we first utilized numerical gait optimization using \textit{GPOPS-II}, and used the obtained gait as an initial guess for our PMP analysis. The resulting gait produces a 45\% greater Lighthill's efficiency than the first generally-shaped efficiency-optimal gait and a 90\% higher Lighthill's efficiency than the large-amplitude efficiency-optimal square-shaped gait.
%V.D - Robotic Swimmer Optimal LHE results - end

\section{Conclusion}
In this work, we presented and analyzed the dynamics of an experimentally realized three-link robotic swimmer operating in drag-dominated, low Reynolds number regime. We found that Purcell's slender links model exhibited qualitative, but not quantitative, agreement with the experimental net displacement of the robotic swimmer. In particular, we found an optimal amplitude that maximizes the net displacement per cycle of a circular gait. By accounting for the added drag of the swimmer's non-slender links and central flotation sphere, we extended Purcell's model with two additional parameters. Calibrating these parameters yielded good agreement with the experimental results. Using an optimal control approach, we obtained generally-shaped periodic gaits that locally maximize the net displacement of the robotic swimmer model under varying joint-angle bounds. Utilizing the differential geometry method developed in \cite{hatton2011geometric} provided a visual interpretation of the variation among these gaits. Finally, we formalized Lighthill's energy-efficiency maximization for both Purcell's slender swimmer model and its robotic model counterpart as an optimal control problem, and obtained newly discovered efficiency-optimal gaits for both models.

We conclude by briefly discussing the limitations of our results and outlining several directions for possible future extensions. First, the optimal gaits identified here have not yet been experimentally demonstrated. Ongoing work aims to implement them experimentally and assess their optimality on an improved robotic swimmer. Second, this work focused on swimmer models with time-invariant dynamics. However, more feasible microswimmer realizations may require time-varying locomotion due to passive-elastic joints and/or magnetic actuation \cite{jang2015undulatory},\cite{gutman2014simple},\cite{PhysRevE.110.014207},\cite{krishnamurthy2017schistosoma}. The optimal control framework developed here can be extended to optimize the motion of such swimmers. Finally, beyond the realm of viscosity-dominated articulated swimmers, future work will extend the proposed optimal control approach to a broader class of locomotion systems, including inertia-dominated swimmers \cite{virozub2019planar} and non-holonomic wheeled vehicles \cite{halvani2022nonholonomic},\cite{rizyaev2026locomotion},\cite{zigelman2026dynamics},\cite{levy2025analysis},\cite{yang2025geometric}.

\useRomanappendicesfalse
\appendices

\section{Proof of constant power in efficiency-optimal gaits} \label{sec:Appendix_A}

\setcounter{equation}{0}
\renewcommand{\theequation}{A\arabic{equation}}
\renewcommand{\theHequation}{A.\arabic{equation}}

In this appendix, we prove Claim \ref{itm:constant-power} from Section \ref{sec:eff_OCP}, which states that a gait with maximal Lighthill’s efficiency must always maintain a constant power expenditure. This proof follows directly from the optimal control formulation.

From (\ref{eq:power}), the time derivative of the mechanical power of a general low-$Re$ three-link swimmer is:
\begin{align} 
    \dot p=& \dot au_1^2 + 2au_1 \dot u_1 +\dot bu_1u_2 + b \dot u_1u_2 + bu_1 \dot u_2 +\nonumber\\ &\dot cu_2^2 + 2cu_2 \dot u_2
    \label{eq:A1}
\end{align}
where:
\begin{align}
    \dot u_1 &= \frac{(\dot b \alpha + b \dot\alpha-2(\dot c\beta+c \dot\beta))\gamma-\dot\gamma(b\alpha-2c\beta)}{\gamma^2} \nonumber\\
    \dot u_2 &= \frac{(\dot b \beta + b \dot\beta-2(\dot a\alpha+a \dot\alpha))\gamma-\dot\gamma(b\beta-2a\alpha)}{\gamma^2} \nonumber\\
    \dot\alpha &= \dot\lambda_2 + \lambda_3\dot h + \dot\lambda_4q + \lambda_4 \dot q \nonumber\\
    \dot\beta &= \dot\lambda_1 + \lambda_3\dot g + \dot\lambda_4f + \lambda_4 \dot f \nonumber\\
    \dot \gamma &= \lambda_5(4\dot ac+4a\dot c-2b\dot b) \nonumber\\
    \dot g &= g_{\phi_1}u_1+g_{\phi_2}u_2+g_{\theta}\dot\theta \nonumber\\
    \dot h &= h_{\phi_1}u_1+h_{\phi_2}u_2+h_{\theta}\dot\theta \nonumber\\
    \dot f &= f_{\phi_1}u_1+f_{\phi_2}u_2 \nonumber\\
    \dot q &= q_{\phi_1}u_1+q_{\phi_2}u_2 \nonumber\\
    \dot a &= a_{\phi_1}u_1+a_{\phi_2}u_2 \nonumber\\
    \dot b &= b_{\phi_1}u_1+b_{\phi_2}u_2 \nonumber\\
    \dot c &= c_{\phi_1}u_1+c_{\phi_2}u_2
    \label{eq:A2}
\end{align}

Substituting equations (\ref{eq:A2}) into equation (\ref{eq:A1}) leads to:
\begin{align}
    \dot p \equiv 0 \rightarrow p(t) \equiv const
    \label{eq:A3}
\end{align}
regardless of the swimmer's state or structure.

\section{Proof of the smoothness of an efficiency-optimal gait at its endpoints}
\label{sec:Appendix_B}

\setcounter{equation}{0}
\renewcommand{\theequation}{B\arabic{equation}}
\renewcommand{\theHequation}{B.\arabic{equation}}

In this appendix, we prove Claim \ref{itm:smoothness} from Section \ref{sec:eff_OCP}, which states that a gait with maximal Lighthill’s efficiency must be smooth at its ends. This proof follows directly from the optimal control formulation.

For a general low-$Re$ three-link swimmer, at the initial time of motion $\phi_1=\phi_2, \theta=0$, and from (\ref{eq:full_connection}),(\ref{eq:power}),(\ref{eq:Trans_Con_LHE}):
\begin{equation}
\left. \begin{bmatrix}
    g=-h\\f=-q\\a=c\\\lambda_1=-\lambda_2
\end{bmatrix} \right|_{t=t_0}
\end{equation}
Substituting into (\ref{eq:Hu_LHE}) one obtains:
\begin{equation}
   \left. \begin{bmatrix} \lambda_1+\lambda_3g+\lambda_4f+\lambda_5(2au_1+bu_2)\\-\lambda_1-\lambda_3g-\lambda_4f+\lambda_5(2au_2+bu_1)\end{bmatrix}\right|_{t=t_0}=\boldsymbol{0}
\end{equation}
Summing the two equations above, one obtains:
\begin{equation}
    (2a+b)(u_1+u_2)|_{t=t_0}=0
\end{equation}
At the initial time, for a symmetric swimmer $(l_1=l_2)$, the expression $2a+b$  is:
\begin{equation}
    2a+b=c_nl^3_1\frac{N_1}{D_1}
    \label{2a_plus_b}
\end{equation}
Where:
\small
\begin{equation}
\begin{array}{l}
N_1=4l_1\sin^2(\phi_1)\!+\!12\nu_1\!+\!2\chi l_0\!+\!(1-\sin^2(\phi_1))\chi l_1 \\ [6pt] D_1=6l_1\sin^2(\phi_1)\!+\!18\nu_1\!+\!3\chi l_0\!+\!(1\!-\!\sin^2(\phi_1))6\chi l_1\\[6pt] \nu_1=\frac{\pi\mu r_s}{c_t}
\end{array}
\end{equation}
\normalsize
Since (\ref{2a_plus_b}) must be positive, the optimal control input must satisfy:
\begin{equation}
   u_1(t_0)=-u_2(t_0)
\end{equation}
regardless of the swimmer's structure or its initial state.

At the final time of motion
$\phi_1\!=\!-\phi_2$, and from (\ref{eq:full_connection}), (\ref{eq:power}), (\ref{eq:Trans_Con_LHE}):
\begin{equation}
\left. \begin{bmatrix} g=h\\a=c\\\lambda_1=\lambda_2\\\lambda_4=0
\end{bmatrix} \right|_{t=t_f}
\end{equation}
Substituting into (\ref{eq:Hu_LHE}) one obtains:
\begin{equation}
   \left. \begin{bmatrix} \lambda_1+\lambda_3g+\lambda_5(2au_1+bu_2)\\\lambda_1+\lambda_3g+\lambda_5(2au_2+bu_1)\end{bmatrix}\right|_{t=t_f}=\boldsymbol{0}
\end{equation}
By subtracting both equations one obtains:
\begin{equation}
    (2a-b)(u_1-u_2)|_{t=t_f}=0
\end{equation}
At the final time point, for a symmetric swimmer $l_1=l_2$, the expression $2a-b$ is:
\begin{equation}
    2a-b=\frac{c_nl^3_1}{3}\frac{N_2}{D_2}
    \label{2a_minus_b}
\end{equation}
Where:
\begin{equation}
\begin{array}{l}
N_2\!=\!192\nu_2\!+\!2\chi\!+\!(1\!-\!z^2)12r\!+\!3\chi rz^2 \\ [6pt] D_2\!=\!96\nu_2\!+\!(1\!-\!z^2)6r\!+\!(6rz^2\!+\!12r^2z\!+\!8z^3\!+\!1) \\[6pt] \nu_2=\frac{\pi\mu r^3_s}{c_tl^3_0}\\[6pt] z=\cos(\phi_1), -1\le z \le 1\\[6pt] r=\frac{l_1}{l_0}>0
\end{array}
\end{equation}
Since (\ref{2a_minus_b}) must be positive, the optimal control input must satisfy:
\begin{equation}
   u_1(t_f)=u_2(t_f)
\end{equation}
regardless of the swimmer's structure or its final state.

\section{Efficiency-suboptimal gaits using truncated Fourier series}
\label{sec:Appendix_C}

\setcounter{equation}{0}
\renewcommand{\theequation}{C\arabic{equation}}
\renewcommand{\theHequation}{C.\arabic{equation}}

This appendix describes how we find sub-optimal gaits and use their final time state values as an initial guess for the optimization of a generally-shaped gait in \ref{sec:LHE_BVP_Sol}.

A relatively simple and straightforward method for finding a sub-optimal gait is by using a gait whose shape is defined by two truncated Fourier series. The final time state of this gait serves as the initial guess for our optimal control solution. We describe the shape of this gait using a geometric variable denoted as $s$:
\begin{equation}
    \boldsymbol{\Phi}(s)=\boldsymbol{R}_{\frac{\pi}{4}}\boldsymbol{\hat\Phi}(s)
    \label{eq:PHI_s}
\end{equation}
Where:
\begin{equation}
    \boldsymbol{\hat\Phi}(s)=\begin{bmatrix}
    \hat\phi_1(s)\\\hat\phi_2(s)
    \end{bmatrix}=\begin{bmatrix} \sum\limits_{n=1}^NA_n\cos((2n-1)s)\\\sum\limits_{n=1}^NB_n\sin((2n-1)s)\end{bmatrix}
\end{equation}
\begin{equation}
    \boldsymbol{R}_{\frac{\pi}{4}}=\begin{bmatrix}
        \cos(\frac{\pi}{4})& -\sin(\frac{\pi}{4})\\ \sin(\frac{\pi}{4})& \cos(\frac{\pi}{4})
\end{bmatrix}
\end{equation}
And $A_n,B_n$ are the scalar coefficients of the truncated Fourier series that are to be optimized. This geometric definition ensures that the examined gaits must be smooth at the edges, as described in appendix B. The rate of change of the joints' angles is therefore:
\begin{equation}
    \boldsymbol{\dot\Phi}(s)=\boldsymbol{R}_{\frac{\pi}{4}}\boldsymbol{\dot{\hat\Phi}}(s)
\end{equation}
Where:
\small
\begin{equation}
    \boldsymbol{\dot{\hat\Phi}}(s)=\begin{bmatrix} \frac{\partial{\hat\phi_1}}{\partial s}\\[3.5pt]\frac{\partial{\hat\phi_2}}{\partial s} \end{bmatrix}\dot s=\begin{bmatrix} -\sum\limits_{n=1}^N(2n-1)A_n\sin((2n-1)s)\\\sum\limits_{n=1}^N(2n-1)B_n\cos((2n-1)s)\end{bmatrix}\dot s
\end{equation}
\normalsize
Since an efficiency-optimal gait must maintain a constant power expenditure, the rate of change $\dot s$ can be determined from (\ref{eq:power}) as a function of $s$:
\begin{equation}
    p_0=P(s)\dot s^2=const\rightarrow \dot s=\sqrt{\frac{p_0}{P(s)}}
\end{equation}
Where:
\begin{equation}
    P(s)=a\left(\frac{\partial{\phi_1}}{\partial s}\right)^2+b\frac{\partial{\phi_1}}{\partial s}\frac{\partial{\phi_2}}{\partial s}+c\left(\frac{\partial{\phi_2}}{\partial s}\right)^2
\end{equation}
Choosing to integrate the state equations about the geometrical variable $s$ one obtains:
\begin{equation}
    \frac{\partial t}{\partial s}=\sqrt{\frac{P(s)}{p_0}}
    \label{eq:dtds}
\end{equation}
And substituting into (\ref{eq:full_connection}):
\begin{equation}
        \frac{\partial x}{\partial s}=g(s,\theta)\frac{\partial \phi_1}{\partial s}+h(s,\theta)\frac{\partial \phi_2}{\partial s}
\end{equation}
\begin{equation}
    \frac{\partial \theta}{\partial s}=f(s)\frac{\partial \phi_1}{\partial s}+q(s)\frac{\partial \phi_2}{\partial s}
    \label{eq:10}
\end{equation}
Arbitrarily defining the constant power expenditure as  $p_0=1$, the optimization problem is fully defined for an unknown final time $t_f$, and Lighthill's efficiency cost function is simplified as:
\begin{equation}
        J(t_f,\textbf z_f)=\frac{x(t_f)}{t_f}
\end{equation}
To find the optimal coefficients $A_n,B_n$, we integrate (\ref{eq:dtds})-(\ref{eq:10}) in the interval of a quarter of a gait $s\in\left[\frac{\pi}{4},\frac{3\pi}{4}\right]$, using Matlab's ode45, and optimize the cost function with the nonlinear algorithm fsolve. We uniformly scale the rate $\dot{s}$ by choosing the constant power $p_0$ that leads to a total quarter-period time of $t_f=1$. Mathematically, this process is formulated as follows. After finding an optimal gait shape, we normalize the shape change rate such that the final time will be $t_f=1$. By integrating (\ref{eq:dtds}) one obtains:
\begin{equation}
    t_f=\frac{1}{\sqrt{p_0}}I(s)
\end{equation}
Where:
\begin{equation}
    I(s)=\int_{\pi/4}^{3\pi/4}\sqrt{P(s)}ds
\end{equation}
Therefore, the proper power expenditure normalization is:
\begin{equation}
    p_0(t_f=1)=t_f^2(p_0=1)
\end{equation}
In this approach, we parameterize a gait as a Fourier series with a constant-power time parametrization to maximize energy efficiency. By optimizing the Fourier coefficients, we obtain an approximate efficiency-optimal gait, which is then used as an initial guess for the PMP-based analysis.

\section*{\textbf{Acknowledgment}}
This work was supported by Israel Science Foundation under grant no. 1382/23. The authors wish to thank Joseph Z. Ben-Asher for useful discussions, and Ross L. Hatton for consultation regarding his software tool \cite{HattonSoftware}.

%\newpage
\bibliographystyle{IEEEtran}
\bibliography{mybibliography}
\end{document}

%% file: my_shortcuts.tex
\newcommand{\vecA}{\mathbf{A}}

\newcommand{\vecD}{\mathbf{D}}

\newcommand{\vecF}{\mathbf{F}}
\newcommand{\vecG}{\mathbf{G}}
\newcommand{\vecH}{\mathbf{H}}

\newcommand{\vecR}{\mathbf{R}}

\newcommand{\vecV}{\mathbf{V}}
\newcommand{\vecW}{\mathbf{W}}

\newcommand{\vecf}{\mathbf{f}}

\newcommand{\vecn}{\mathbf{n}}

\newcommand{\vecq}{\mathbf{q}}

\newcommand{\vect}{\mathbf{t}}
\newcommand{\vecu}{\mathbf{u}}
\newcommand{\vecv}{\mathbf{v}}

\newcommand{\vecz}{\mathbf{z}}

\newcommand{\veclambda}{\boldsymbol\lambda}